\documentclass[11pt,a4paper]{article}
\usepackage{lscape}
\usepackage{color}
\usepackage{colortbl}
\usepackage{amsmath}
\usepackage{amssymb}
\usepackage{amsfonts}
\usepackage{float}
\usepackage{pifont}
\usepackage{multirow}
\usepackage{graphics}
\usepackage{fullpage}
\newcommand{\remark}[1]{\noindent {\bf Remark.} #1 \par}

\newcommand{\itemain}{$\bullet$}
\newcommand{\hashfam}{\ensuremath{\mathfrak{h}}}
\newcommand{\descprot}{Description of the protocol}
\newcommand{\algo}{Algorithm}
\newcommand{\entree}{Input:}
\newcommand{\sortie}{Output:}
\newcommand{\sialorssinonl}[3]{\textsf{if} #1 \textsf{then} #2 \\
\phantom{\textbf{if} #1} \textsf{else} #3}
\newcommand{\sialorssinon}[3]{\textsf{if}  #1 \textsf{then}  #2 \textsf{else} #3}

\newcommand{\Exp}{\textsf{Exp}}
\newcounter{Jeuxcount}
\newenvironment{Jeux}%
{\begin{list}{\Exp$_{\arabic{Jeuxcount}}$} {\usecounter{Jeuxcount}} \rm}
{\end{list}}

\newtheorem{definition}{Definition}
\newtheorem{theorem}{Theorem}

\newcommand{\COKEA}          {KEA}

\newcommand{\fonction}[2]{\ensuremath{\mathcal{F}(#1,#2)}}

\newcommand{\eps}{\ensuremath{\varepsilon}}
\newcommand{\llb}{[\![}
\newcommand{\rrb}{]\!]}
\newcommand{\range}[1]{\llb 1 , #1 \rrb}
\newcommand{\lra}{\ensuremath{\longrightarrow}}

\newcommand{\HList}{\ensuremath{\mathsf{H\hbox{-}List}}}

\newcommand{\GG}{\mathbb{G}}

\newcommand{\NN}{\mathbb{N}}
\newcommand{\ZZ}{\mathbb{Z}}

\newcommand{\hasard}{\xleftarrow{R}}
\newcommand{\couple}[2]{\ensuremath{\langle #1 , #2 \rangle}}
\newcommand{\Pa}{\ensuremath{P_1}}
\newcommand{\GGa}{\ensuremath{\GG_1}}
\newcommand{\Pb}{\ensuremath{P_2}}
\newcommand{\GGb}{\ensuremath{\GG_2}}

\newcommand{\GGc}{\ensuremath{\GG_3}}
\newcommand{\expo}[2]{\ensuremath{[#1] #2}}
\newcommand{\expocouple}[3]{\expo{#1}{\couple{#2}{#3}}}
\newcommand{\bilinear}{(q,\GGa,\GGb,\GGc,\couple{\cdot}{\cdot},\psimorph)}
\newcommand{\psimorph}{\psi}

\newcommand{\dvsbbprobcourt}[1]   {\ensuremath{\mathcal{PR}_1(#1)}}
\newcommand{\sigst}    {\ensuremath{\sigma^{\star}}}
\newcommand{\dsigst}   {\ensuremath{\tau^{\star}}}
\newcommand{\COLL}{\ensuremath{\mathsf{Collision}}}

\newcommand{\pks}{\ensuremath{\mathsf{pk_s}}}
\newcommand{\sks}{\ensuremath{\mathsf{sk_s}}}
\newcommand{\pkc}{\ensuremath{\mathsf{pk_c}}}
\newcommand{\skc}{\ensuremath{\mathsf{sk_c}}}

\newcommand{\mst}      {\ensuremath{m^{\star}}}

\newcommand{\etali}{\emph{et al.}}

\newcommand{\ie}{\emph{i.e.}}
\newcommand{\eg}{\emph{e.g.}}
\newcommand{\resp}{\emph{resp.}}

\newcommand{\Hash}    {\ensuremath{\mathcal{H}}}
\newcommand{\HashFam} {\ensuremath{\mathfrak{H}}}

\newcommand{\qHash}   {\ensuremath{q_{\HashFam}}}
\newcommand{\orHash}  {\ensuremath{\mathcal{O}_{\HashFam}}}

\newcommand{\hash}{\ensuremath{{h}}}

\newcommand{\gen}{\ensuremath{\mathsf{Gen}}}

\newcommand{\Adver}{\ensuremath{\mathcal{A}}}

\newcommand{\AdverB}{\ensuremath{\mathcal{B}}}
\newcommand{\AdverC}{\ensuremath{\mathcal{C}}}

\newcommand{\Exper}[3]   {\ensuremath{\mathbf{Exp}_{#1,#2}^{#3}}}
\newcommand{\Succ}[3]{\ensuremath{\mathbf{Succ}_{#1,#2}^{#3}}}
\newcommand{\Adv}[3]{\ensuremath{\mathbf{Avan}_{{#1},{#2}}^{#3}}}
\newcommand{\orSig}{\ensuremath{\mathfrak{S}}}
\newcommand{\orCon}{\ensuremath{\mathfrak{V}}}

\newcommand{\CDH}{{CDH}}

\newcommand{\DDH}{{DDH}}

\newcommand{\DBDH}{{DBDH}}

\newcommand{\bbprob}[1]{\ensuremath{#1\hbox{-}\mathsf{SDH}}}

\newcommand{\bls}{\ensuremath{\mathsf{BLS}}}

\newcommand{\bobo}{\ensuremath{\mathsf{BB}}}

\newcommand{\blsudvs}{\ensuremath{\mathsf{UDVS\hbox{-}BLS}}}

\newcommand{\bbudvs}{\ensuremath{\mathsf{UDVS\hbox{-}BB}}}
\newcommand{\blsumdvs}{\ensuremath{\mathsf{UDVS\hbox{-}BLS(n)}}}

\newcommand{\setup}{\textsf{Setup}}
\newcommand{\sign}{\textsf{Sign}}

\newcommand{\designate}{\textsf{Designate}}
\newcommand{\verify}{\textsf{Verify}}
\newcommand{\dverify}{\textsf{DVerify}}
\newcommand{\keygen}{\textsf{KeyGen}}
\newcommand{\skeygen}{\textsf{SKeyGen}}
\newcommand{\vkeygen}{\textsf{VKeyGen}}

\newcommand{\fake}{\textsf{Fake}}
\newcommand{\sig}{\sigma}
\newcommand{\dsig}{\tau}

\newcommand{\valid}{1}
\newcommand{\invalid}{0}
\newcommand{\params}{\ensuremath{\mathcal{P}}}

\newcommand{\pkS}{\ensuremath{\mathsf{pk_s}}}
\newcommand{\skS}{\ensuremath{\mathsf{sk_s}}}
\newcommand{\pkV}{\ensuremath{\mathsf{pk_v}}}
\newcommand{\skV}{\ensuremath{\mathsf{sk_v}}}

\newcommand{\probleme}{\ensuremath{\mathcal{PR}_i}}

\newcommand{\dbdh}{\DBDH} 
 
\newcommand{\flex}{\ensuremath{\mathcal{PR}_2}}
\newcommand{\sfflex}{\ensuremath{\textsf{\flex}}}
\newcommand{\dsfflex}{\ensuremath{\mathcal{PR}_3}}

\newcommand{\sfkea}{\ensuremath{\textsf{KEA}}}

\newcommand{\qSig}{q_{\mathfrak{S}}}
\newcommand{\qCon}{q_{\mathfrak{V}}}
\newcommand{\UDVSEFCMA}  {\ensuremath{\mathsf{UDVS}\hbox{-}\mathsf{EF}\hbox{\textsf{-}}\mathsf{CMA}}}

\newcommand{\USPSICMA}   {\ensuremath{\mathsf{UDVS}\hbox{-}{\Psi}\hbox{\textsf{-}}\mathsf{CMA}}}

\newcommand{\EFCMA}      {\ensuremath{\mathsf{EF}\hbox{\textsf{-}}\mathsf{CMA}}}

\newcommand{\bfA}{\ensuremath{\mathcal{A}}}
\newcommand{\bfAbar}{\ensuremath{\overline{\mathcal{A}}}}
\newcommand{\Texp}[1]{\ensuremath{T_{\hbox{exp}}(#1)}}

\title{New Extensions of Pairing-based Signatures into \\ Universal (Multi) Designated Verifier Signatures\footnote{\noindent This is
      the full version of ``New Extensions of Pairing-based Signatures
      into Universal Designated Verifier Signatures''
      \cite{Vergnaud2006a} presented at ICALP'06.}}

\author{Damien Vergnaud}

\date{}

\begin{document}

\maketitle

\vspace{-.8cm}
\begin{center}
\'Ecole Normale Sup\'erieure -- C.N.R.S. -- I.N.R.I.A. \\
45 rue d'Ulm, 75230 Paris Cedex 05 -- France
\end{center}

\vspace{.8cm}
\begin{abstract}
\noindent  The concept of universal designated verifier signatures was
  introduced by Steinfeld, Bull, Wang and Pieprzyk at Asiacrypt 2003.
  These signatures can be used as standard publicly verifiable digital
  signatures but have an additional functionality which allows any
  holder of a signature to designate the signature to any desired
  verifier. This designated verifier can check that the message was
  indeed signed, but is unable to convince anyone else of this fact.
  We propose new efficient constructions for pairing-based short
  signatures. Our first scheme is based on Boneh-Boyen signatures and
  its security can be analyzed in the standard security model. We
  prove its resistance to forgery assuming the hardness of the so-called
  strong Diffie-Hellman problem, under the knowledge-of-exponent
  assumption. The second scheme is compatible with the
  Boneh-Lynn-Shacham signatures and is proven unforgeable, in the
  random oracle model, under the assumption that the computational
  bilinear Diffie-Hellman problem is untractable. Both schemes are
  designed for devices with constrained computation capabilities since
  the signing and the designation procedure are pairing-free. Finally,
  we present extensions of these schemes in the multi-user setting
  proposed by Desmedt in 2003.

\end{abstract}

\section{Introduction} Recently many \emph{universal designated
  verifier signature} protocols have been proposed (\eg{}
\cite{LaguillaumieVergnaud2005a,SteinfeldBullWangPieprzyk2003,ZhangFurukawaImai2005}).
The present paper focuses on the proposal of two new efficient
constructions for pairing-based short signatures
\cite{BonehBoyen2004,BonehLynnShacham2004} and on the security
treatment of them. The resistance to forgery of the first scheme
relies on the hardness of the strong Diffie-Hellman problem, under the
\emph{knowledge-of-exponent assumption}, in the standard security
model, and the one of the second scheme relies, in the random oracle
model, on the hardness of a new computational problem (not easier than
the widely used computational bilinear Diffie-Hellman problem).

\subsection{Related work.} Many cryptographic primitives have been
proposed to limit the \textit{self-authen\-ti\-cating} property of
digital signatures. The primary one: \emph{undeniable signatures} --
introduced by Chaum and van Antwerpen in 1989
\cite{ChaumvanAntwerpen1990} -- appeared to have some weaknesses.  The
concept of \emph{designated verifier signatures} was introduced by
Jakobsson, Sako and Impagliazzo \cite{JakobssonSakoImpagliazzo1996} in
order to repair their so-called \emph{lie detector problem}.
Designated verifier signatures are intended to a specific and unique
designated verifier, who is the only one able to check their validity.
Motivated by privacy issues associated with dissemination of signed
digital certificates, Steinfeld, Bull, Wang and Pieprzyk
\cite{SteinfeldBullWangPieprzyk2003} defined, in 2003, a new kind of
signatures called \emph{universal designated-verifier signatures}
(UDVS). This primitive can function as a standard publicly-verifiable
digital signature scheme but has an additional functionality which
allows any holder of a signature to designate the signature to any
verifier. Again, the designated-verifier can check that the message
was signed by the signer, but is unable to convince anyone else of
this fact. Designated verifier signatures (universal or not) have
found numerous applications in financial cryptography (\emph{e.g.}
call for tenders, electronic voting, electronic auction or distributed
contract signing).

Steinfeld \emph{et al.} proposed an efficient UDVS scheme constructed
using any bilinear group-pair. In collaboration with Laguillaumie, we
suggested in \cite{LaguillaumieVergnaud2005a} a variant which
significantly improves this protocol. Both schemes are compatible with
the key-generation, signing and verifying algorithms of the
Boneh-Lynn-Shacham \cite{BonehLynnShacham2004} signature scheme
(\bls). In \cite{BonehBoyen2004}, Boneh and Boyen proposed efficient
pairing-based short signatures (\bobo) whose security can be analyzed
in the standard security model. A UDVS scheme compatible with a
variant of Boneh and Boyen's scheme has been proposed by Zhang,
Furukawa and Imai \cite{ZhangFurukawaImai2005}.

\subsection{Contributions of the paper.} The main contribution of the
paper is to provide a new efficient UDVS protocol compatible with the
\emph{original} Boneh-Boyen scheme. The idea underlying our design
relies on the flexibility of \bobo\ signatures and specifically on
their good behaviour under scalar multiplication.  The new scheme,
that we call \bbudvs, is unforgeable in the standard security model
assuming the hardness of the strong Diffie-Hellman problem
\cite{BonehBoyen2004}, under the knowledge-of-exponent assumption
(KEA) \cite{BellarePalacio2004,Damgard1991}. The protocol proposed by
Zhang \etali{} is proven unforgeable assuming the hardness of the same
algorithmic problem, but under an additional assumption (which is
stronger than KEA). The security of \bbudvs{} can
also be proved under a well-defined (though \emph{ad hoc}) computational
problem without using any non-black-box assumption (such as KEA). The
computational workload of \bbudvs{} amounts to three exponentiations
over bilinear groups for designating a signature and four pairing
evaluations to verify it, and moreover, the length of the signatures
is much smaller than the one of Zhang \etali{}'s signatures. Following
the general paradigm from \cite{LaguillaumieVergnaud2004}, this scheme
is readily extended to produce universal multi designated verifier
signatures (UMDVS) \cite{NgSusiloMu2005} that are verifiable in a
non-interactive way. The multi-user scheme inherits the efficiency
properties of \bbudvs{} with the same signature size (which, in
particular, does not grow with the number of verifiers).

Using the same design principle, we propose a new UDVS protocol
compatible with the \bls\ signatures which is well-suited for devices
with constrained computation capabilities and low bandwidth.  Indeed
the designation procedure of the signatures is pairing-free and the
resulting size is comparable to the length of DSA signatures. The
proof of security for this scheme, that we called \blsudvs, takes
place in the random oracle model \cite{BellareRogaway1993}: we show
that this scheme is unforgeable with respect to a new computational
assumption weaker than the widely used computational bilinear
Diffie-Hellman assumption. In some cases
\cite{JakobssonSakoImpagliazzo1996,LaguillaumieVergnaud2005a} it may
be desirable that UDVSs provide a stronger notion of privacy. The
scheme \blsudvs{} provides this security requirement assuming the
hardness of the $xyz$-decisional co-Diffie Hellman problem.  It is
possible to extend this scheme into a UMDVS one.

\section{Definitions}
\subsection{Notations}
The set of $n$-bit strings is denoted by $\{0,1\}^n$ and the set of
all finite binary strings is denoted by $\{0,1\}^*$. Let $\mathcal{A}$
be a probabilistic {Turing} machine running in polynomial time (a
PPTM, for short), and let $x$ be an input for $\mathcal{A}$. The
probability space that assigns to a string $\sigma$ the probability
that $\Adver$, on input $x$, outputs $\sigma$ is denoted by
$\Adver(x)$. The support of $\Adver(x)$ is denoted by $\Adver[x]$.
Given a probability space $S$, a PPTM that samples a random element
according to $S$ is denoted by $x \xleftarrow{R} S$. For a finite set
$X$, $x \xleftarrow{R} X$ denotes a PPTM that samples a random element
uniformly at random from $X$.

\subsection{Universal designated verifier signatures}
In this subsection, we recall the definitions of UDVS schemes and of
their security requirements
\cite{LaguillaumieVergnaud2004,SteinfeldBullWangPieprzyk2003}.

\subsubsection{Syntactic definition}

\begin{definition} A \emph{universal designated verifier signature
    scheme} $\Sigma$ is an 8-tuple 
$$\Sigma =
  (\setup,\skeygen,\vkeygen,\sign,\verify, \designate,\fake,\dverify)$$
  such that \begin{itemize} \item $(\setup,\skeygen,\sign,\verify)$ is
    a signature scheme: \begin{itemize} \item $\Sigma.\setup$ is a
      PPTM which takes an integer $k$ as input. The output are the
      \emph{public parameters} $\params$. $k$ is called the
      \emph{security parameter}.\item $\Sigma.\skeygen$ is a PPTM
      which takes the public parameters as input. The output is a pair
      $(\skS,\pkS)$ where $\skS$ is called a \emph{signing secret key}
      and $\pkS$ a \emph{signing public key}. \item $\Sigma.\sign$ is
      a PPTM which takes the public parameters, a message, and a
      signing secret key as inputs and outputs a bit string. \item
      $\Sigma.\verify$ is a PPTM which takes the public parameters, a
      message $m$, a bit string $\sig$ and a signing public key
      $\pkS$. It outputs a bit. If the bit output is \valid\ then the
      bit string $\sig$ is said to be a \emph{signature} on $m$ for
      $\pkS$.
    \end{itemize} \item $\Sigma.\vkeygen$ is a PPTM which takes the
    public parameters as input. The output is a pair $(\skV,\pkV)$
    where $\skV$ is called a \emph{verifying secret key} and $\pkV$ a
    \emph{verifying public key}. \item $\Sigma.\designate$ is a PPTM
    which takes the public parameters, a message $m$, a signing public
    key $\pkS$, a signature $\sig$ on $m$ for $\pkS$ and a verifying
    public key as inputs and outputs a bit string. \item
    $\Sigma.\fake$ is a PPTM which takes the public parameters, a
    message, a signing public key and a verifying secret key as inputs
    and outputs a bit string. \item $\Sigma.\dverify$ is a
    deterministic PPTM which takes the public parameters, a message
    $m$, a bit string $\tau$, a signing public key $\pkS$, a verifying
    public key $\pkV$ and the matching verifying secret key $\skV$ as
    inputs. It outputs a bit. If the bit output is \valid\ then $\tau$
    is said to be a \emph{designated verifier signature} on $m$ from
    $\pkS$ to $\pkV$.
  \end{itemize} $\Sigma$ must satisfies the following properties, for
  all $k \in \NN$, all $\params \in \Sigma.\setup[k]$, all
  $(\pkS,\skS) \in \Sigma.\skeygen[\params]$, all $(\pkV,\skV) \in
  \Sigma.\vkeygen[\params]$ and all messages $m$: \begin{itemize}
  \item \textsc{Correctness of Signature:} $$\forall \sig \in
    \Sigma.\sign[\params,m,\skS], \ \
    \Sigma.\verify[\params,m,\sig,\pkS] = \{\valid\}. $$ \item
    \textsc{Correctness of Designation:} $$\begin{array}{c} \forall
      \sig \in \Sigma.\sign[\params,m,\skS], \forall \tau \in
      \Sigma.\designate[\params, m, \pkS, \sig, \pkV], \\
      \Sigma.\dverify[\params,m,\tau,\pkS, \pkV, \skV] = \{\valid\}.
    \end{array}	$$
  \item \textsc{Source Hiding:}
$$
\Sigma.\designate(\params,m,\pkS,\Sigma.\sign(\params,m,\skS),\pkV]) =
\Sigma.\fake(\params,m,\pkS,\skV).
$$
\end{itemize}
\end{definition}

\noindent The correctness properties insure that a properly formed
(designated verifier) signature is always accepted by the (designated)
verifying algorithm. The source hiding property states that given a
message $m$, a signing public key $\pkS$, a verifying public key
$\pkV$ and a designated verifier signature $\tau$ on $m$ from $\pkS$
to $\pkV$ it is (unconditionally) infeasible to determine if $\tau$
was produced by $\Sigma.\designate$ or $\Sigma.\fake$.

\subsubsection{Security requirements}
In this section, we state the definitions of \emph{unforgeability} and
\emph{privacy of signer's identity} under a chosen message attack that
were introduced in
\cite{LaguillaumieVergnaud2005a,SteinfeldBullWangPieprzyk2003}. 

In the
following $\Sigma =
(\setup,\skeygen,\vkeygen,\sign,\verify,\designate,\fake,\dverify)$
denotes a UDVS scheme.

\bigskip
\paragraph{Resistance to forgery.}
The accepted definition of security for signature schemes is
existential unforgeability under adaptive chosen message attack
\cite{GoldwasserMicaliRivest1988}. The notion of \UDVSEFCMA-security
\cite{LaguillaumieVergnaud2005a,SteinfeldBullWangPieprzyk2003} is a
natural extension of this to the UDVS setting.

It is defined \emph{via} a random experiment parameterized by a
security parameter $k$. The experiment involves an adversarial user
$\mathcal{A}$ and is as follows: first two public/secret key pairs for
the signer and the verifier are generated by running the key
generation algorithms.  Then $\mathcal{A}$ engages in polynomially
many runs of the signing oracle, the verifying oracle and -- possibly
-- the random oracle, interleaved at its own choosing. Eventually,
$\Adver$ outputs a pair $(m^{\star},\tau^{\star})$, such that
$m^{\star}$ was never queried to the signing oracle, and it wins if
the verifying oracle returns \valid{} when queried on this pair.

\medskip
\begin{definition}\label{def:UDVS-EFCMA}
  Let $\Adver$ be a PPTM. We consider the following random
  experiments, where $k \in \NN$ is a security parameter:
  \begin{center}
    \begin{tabular}{l}
      \fbox{{Experiment} $\mathbf{Exp}_{\Sigma,
          \Adver}^{\UDVSEFCMA}(k)$}\\
      $\mathcal{L} \leftarrow \emptyset$ \\
      $\params \hasard \Sigma.\setup(k)$ \\
      $(\skS, \pkS) \hasard \Sigma.\skeygen(\params)$ ;
      $(\skV, \pkV) \hasard \Sigma.\vkeygen(\params)$ \\

      $(m^{\star},\tau^{\star}) \hasard
      \Adver^{\orSig,\orCon}(\params,\pkS,\pkV)$ \\
      \phantom{$(m^{\star},\tau^{\star}) \hasard$} $\left\vert
        \begin{array}{l}
          \orSig:  m \dashrightarrow
          \Sigma.\sign(\params, m, \skS)  ;  \mathcal{L} \leftarrow
          \mathcal{L} \cup \{m\} \\
          \orCon: (m, \tau) \dashrightarrow \Sigma.\dverify(\params, m, \tau, \pkS, \pkV, \skV) \end{array} \right.$ \\

      \textbf{return} $1$ \textbf{if} $\Sigma.\dverify(\params,
      m^{\star},\tau^{\star},\pkS,\pkV, \skV) = \{\valid\}$ \textbf{and} $m^{\star} \notin \mathcal{L}$ \\
      \phantom{{\textbf{return}}} $0$ \textbf{otherwise}.
    \end{tabular}
  \end{center}

  \smallskip
  \noindent Let $\tau,\qSig,\qCon \in {\NN}^{\NN}$, $\eps \in
  [0,1]^{\NN}$. We define the \emph{success} of $\mathcal{A}$
  \emph{via} $$ \mathbf{Succ}_{\Sigma,\mathcal{A}}^{\UDVSEFCMA}(k) =
  \Pr[\mathbf{Exp}_{\Sigma,\mathcal{A}}^{\UDVSEFCMA}(k)=1]. $$

  \begin{enumerate}
  \item $\Adver$ is a \emph{$(\tau,\qSig,\qCon)$-\UDVSEFCMA-adversary}
    if for all $k \in \mathbb{N}$, the experiment
    $\mathbf{Exp}_{\Sigma,\mathcal{A}}^{\UDVSEFCMA}(k)$ ends in
    expected time less than $\tau(k)$ and in this experiment
    $\mathcal{A}$ makes at most $\qSig(k)$ (\resp $\qCon(k)$) queries
    to the oracle $\orSig$ (\resp $\orCon$).
  \item $\Sigma$ is \emph{$(\tau,\qSig,\qCon,\eps)$-\UDVSEFCMA-secure}
    if for any {$(\tau,\qSig,\qCon)$-\UDVSEFCMA-adversary} $\Adver$
    and any $k \in \mathbb{N}$,
    $\mathbf{Succ}_{\Sigma,\Adver}^{\UDVSEFCMA}(k) \leq \eps(k)$.
  \end{enumerate}
\end{definition}

\smallskip
\noindent This definition does not capture that the adversary cannot
generate a new signature on a previously signed message (the so-called
\emph{strong unforgeability}).

\bigskip
\paragraph{Privacy of signer's identity.}
As explained in \cite{JakobssonSakoImpagliazzo1996}, in some cases, it
may be desirable that designated verifier signatures provide a
stronger notion of privacy. More precisely, given a designated
verifier signature and two potential signing public keys, it should be
computationally infeasible for an eavesdropper, to determine under
which of the two corresponding secret keys the signature was
performed. The \emph{privacy of signer's identity} ($\Psi$) property
was formalized in \cite{LaguillaumieVergnaud2005a} to capture this
security notion.

We consider a \USPSICMA-adversary ${\mathcal A}$, which runs in two
stages: in the \texttt{find} stage, it takes two signing public keys
$\pkS_0$ and $\pkS_1$ and a verifying public key $\pkV$, and outputs a
message $m^{\star}$ together with some state information
$\mathcal{I}^{\star}$. In the \texttt{guess} stage, it gets a
challenge UDVS $\tau^{\star}$ formed at random under one of the two
keys and the information $\mathcal{I}^{\star}$, and must say which key
was chosen. The adversary has access to the signing oracles $\orSig$,
to the verifying oracle $\orCon$ and -- possibly -- to a random
oracle. It is allowed to invoke them on any message with the
restriction of not querying $m^{\star}$ from $\orSig$ or $\orCon$ in
any stage.

\medskip
\begin{definition}
  Let $\Adver$ be a PPTM. We consider the following random
  experiments, where $b \in \{0,1\}$ and $k \in \NN$ is a security
  parameter:
  \begin{center}
    \begin{tabular}{l}
      \fbox{{Experiment} $\mathbf{Exp}_{\Sigma, \Adver}^{\USPSICMA-b}(k)$}\\
      $\mathcal{L} \leftarrow
      \emptyset$ \\
      $\params \hasard \Sigma.\setup(k)$ \\ 

      $(\skS_0, \pkS_0) \hasard \Sigma.\skeygen(\params)$ ; $(\skS_1,
      \pkS_1) \hasard \Sigma.\skeygen(\params)$, \\
      $(\skV, \pkV) \hasard \Sigma.\vkeygen(\params)$ \\

      $(m^{\star},\mathcal{I}^{\star}) \hasard
      \Adver^{\orSig,\orCon}(\texttt{find},\params,\pkS_0,\pkS_1,\pkV)$
      \\

      \phantom{---} $\left\vert
        \begin{array}{l} \orSig:  (m,i) \dashrightarrow
          \Sigma.\sign(\params, m, \skS_i)  ;  \mathcal{L} \leftarrow
          \mathcal{L} \cup \{m\}
          \\ \orCon: (m, \tau,i) \dashrightarrow
          \Sigma.\dverify(\params, m, \tau, \pkS_i, \pkV, \skV)  ; \mathcal{L} \leftarrow
          \mathcal{L} \cup \{m\}
        \end{array} \right.$ \\
      $\sigma^{\star} \hasard \Sigma.\sign(\params,m^{\star},\skS_{b})$ ;
      $\tau^{\star} \hasard
      \Sigma.\designate(\params,m^{\star},\pkS_{b},\sigma^{\star},\pkV)$
      \\

      $b^{\star} \hasard
      \Adver^{\orSig,\orCon}(\texttt{guess},\tau^{\star},\mathcal{I}^{\star})$
      \\

      \textbf{return} $1$ \textbf{if} $b = b^{\star}$ \textbf{and}
      $m^{\star} \notin \mathcal{L}$ \\
      \phantom{\textbf{return}} $0$ \textbf{otherwise}.
    \end{tabular}
  \end{center}

  \smallskip
  \noindent Let $\tau,\qSig,\qCon \in {\NN}^{\NN}$, $\eps \in
  [0,1]^{\NN}$. We define the \emph{advantage} of $\mathcal{A}$
  \emph{via}
$$
\mathbf{Adv}_{\Sigma,\mathcal{A}}^{\USPSICMA}(k) = \Big\vert
\Pr[\mathbf{Exp}_{\Sigma,\mathcal{A}}^{\USPSICMA-0}(k)=1] -
\Pr[\mathbf{Exp}_{\Sigma,\mathcal{A}}^{\USPSICMA-1}(k)=1] \Big\vert.
$$

\begin{enumerate}
\item $\Adver$ is a \emph{$(\tau,\qSig,\qCon)$-\USPSICMA-adversary} if
  for all $k \in \mathbb{N}$, the experiment
  $\mathbf{Exp}_{\Sigma,\mathcal{A}}^{\USPSICMA}(k)$ ends in expected
  time less than $\tau(k)$ and in this experiment $\mathcal{A}$ makes
  at most $\qSig(k)$ (\resp $\qCon(k)$) queries to the oracle $\orSig$
  (\resp. $\orCon$).
\item $\Sigma$ is \emph{$(\tau,\qSig,\qCon,\eps)$-\USPSICMA-secure} if
  for any {$(\tau,\qSig,\qCon)$-\USPSICMA-adversary} $\Adver$ and any
  $k \in \mathbb{N}$, $\mathbf{Adv}_{\Sigma,\Adver}^{\USPSICMA}(k)
  \leq \eps(k)$.
\end{enumerate}
\end{definition}

\bigskip \remark{Recently, Lipmaa, Wang and Bao
  \cite{LipmaaWangBao2005} have identified a new security requirement
  for designated verifier signatures, that they called the
  \emph{non-delegatability}. This property captures the infeasibility
  for a signer to delegate her authentication capacity without
  revealing her private key. In spite of its interest, we do not
  consider this issue in the following.  Indeed, in this paper, we
  focus on UDVSs, and it is quite easy to see that, if the underlying
  designated verifier signature scheme is non-delegatable then the
  basic signature scheme is universally forgeable under a
  chosen-message attack. In \cite{Vergnaud2006c}, we propose a new
  definition of non-delegatability for UDVS schemes and present some
  new schemes achieving this security requirement.}

\subsection{Bilinear maps and computational assumptions}\label{sec:pbs}
The security of asymmetric cryptographic tools relies on assumptions
about the hardness of certain algorithmic problems.  Bilinear maps
such as Weil or Tate pairing on elliptic curves and hyperelliptic
curves have found various applications in cryptography (\eg\
\cite{BonehBoyen2004,BonehLynnShacham2004}). In the following, we
review the definition of cryptographic bilinear maps and in order to
highlight that our schemes apply to any instantiation of \bls\ and
\bobo\ signatures, we do not pin down any particular generator, but
instead parameterize definitions and security results by a choice of
generator.

\medskip
\begin{definition} A \emph{prime-order-BDH-para\-meter-generator} is a
  PPTM that takes as input $k \in \mathbb{N}$ and outputs a tuple
  $\bilinear$ satisfying the following conditions:
  \begin{enumerate}
  \item $q$ is a prime with $2^{k-1} < q < 2^k$;
  \item $(\GG_1,+)$, $(\GG_2, +)$ and $(\GGc,\cdot)$ are groups of
    order $q$;
  \item $\psimorph: \GG_2 \lra \GG_1$ is an isomorphism s.t. there
    exists a PPTM to compute $\psimorph$;
  \item $\couple{\cdot}{\cdot} : \GG_1 \times \GG_2 \longrightarrow
    \GGc$ satisfies the following properties:
    \begin{enumerate}
    \item $\couple{\expo{a}{Q}}{\expo{b}{R}}=\couple{Q}{R}^{ab}$ for
      all $(Q,R) \in \GG_1 \times \GG_2$ and all $(a,b) \in \ZZ^2$;
    \item $\couple{\cdot}{\cdot}$ is non degenerate (\ie\
      $\couple{\psimorph(P)}{P} \neq 1_{\GGc}$ for some $P \in
      \GG_2$);
    \item there exists a PPTM to compute $\couple{\cdot}{\cdot}$.
    \end{enumerate}
  \end{enumerate}
\end{definition}

\medskip Let $\bilinear$ be as above, $P_2 \in \GG_2$ and let $P_1 =
\psimorph(P_2)$. In margin to the classical Diffie-Hellman problems in
the groups $\GG_1$, $\GG_2$ and $\GGc$, the introduction of bilinear
maps in cryptography gives rise to new algorithmic problems
\cite{BonehLynnShacham2004,LaguillaumieVergnaud2005a}. For instance,
to analyze the security of their signatures, Boneh and Boyen
\cite{BonehBoyen2004} introduced a new computational problem, on which
relies also the unforgeability of our scheme \bbudvs:

\medskip\noindent \textsf{$\ell$-Strong Diffie-Hellman
  (\bbprob{\ell}): } let $x \in \llb 1,q-1\rrb$. Given $\ell \in \mathbb{N}$ and $(\expo{x}{P_2},\ldots,\expo{x^{\ell}}{P_2}) \in
\GG_2^{\ell},$ compute a pair $\left(\expo{(x+m)^{-1}}{P_1} ,m\right)$
in $\GG_1 \times \range{q-1}$.

\medskip
\noindent We will prove the unforgeability of \bbudvs{} assuming the
intractability of this problem under KEA and the one of a new ad-hoc
problem (but not easier than the previous one under KEA):

\medskip\noindent \textsf{\dvsbbprobcourt{\ell}: } let $x,y$ be two
integers smaller than $q$. Given $\ell \in \NN$,
$(m_1,\ldots,m_{\ell}) \in \range{q}^{\ell}$ and
$$(\expo{(x+m_1)^{-1}}{P_2},\ldots,\expo{{(x+m_\ell)^{-1}}}{P_2}) \in
\GG_2^{\ell},$$ compute a $4$-tuple $(m,R,S,T)$ in $\left(\range{q-1}
  \setminus \{m_1,\ldots,m_{\qSig(k)}\}\right) \times \GGa^3$ such
that
\begin{equation}\label{eq:DVSBBpreuve}
  \couple{S}{X+\expo{m}{\Pb}} = \couple{R}{\Pb} \hbox{ and }
  \couple{T}{\Pb} = \couple{R}{Y}.
\end{equation}

\medskip The unforgeability of \blsudvs{} relies also on a new
algorithmic problem (but not easier than the widely used computational
bilinear Diffie-Hellman problem):

\medskip\noindent \sfflex: let $x$, $y$, $z$ be three integers smaller
than $q$. Given $\expo{x}{P_1}$, $\expo{y}{P_2}$ and $\expo{z}{P_2}$,
compute a pair $(R,Q) \in \GG_1 \times \GG_2$ such that
$\couple{R}{Q}=\couple{P_1}{P_2}^{xyz}$.

\medskip\noindent Its \USPSICMA-security relies on the decisional
variant of it that we denote \dsfflex.

\medskip
\begin{definition} Let $\ell \in {\NN}^{\NN}$ and $\Adver$ be a PPTM.
  We consider the following random experiments, where $k \in \NN$ is a
  security parameter:

  \smallskip
  \begin{center}
    \noindent\begin{tabular}{ll}
      \fbox{Experiment $\mathbf{Exp}_{\gen,\Adver}^{{\dvsbbprobcourt{\ell}}}(k)$} & \fbox{Experiment $\mathbf{Exp}_{\gen,\Adver}^{\sfflex}(k)$}  \\

      $\params = \bilinear \xleftarrow{R} \gen(k)$ & $\params = \bilinear \xleftarrow{R} \gen(k)$ \\

      $P_2 \xleftarrow{R} \GG_2 \setminus \{\mathbb{O}_{\GG_2}\}$ ; $(x,y) \xleftarrow{R} \range{q-1}^2$  & $P_2 \xleftarrow{R} \GG_2 \setminus \{\mathbb{O}_{\GG_2}\}$  ; $(x,y,z) \xleftarrow{R} \range{q-1}^3$  \\

      $X \leftarrow \expo{x}{P_2}, Y \leftarrow \expo{y}{P_2}$ & $X \leftarrow \expo{x}{\psimorph(P_2)}, Y \leftarrow \expo{y}{P_2}, Z \leftarrow \expo{z}{P_2}$ \\ 

      \textbf{for} $i$ \textbf{from} $1$ \textbf{to} $\ell(k)$ \textbf{do} & $(R,Q)
      \xleftarrow{R} \mathcal{A}(\params,P_2,X,Y,Z)$ \\

      \phantom{--} $m_i \hasard \range{q}$ ; $R_i \leftarrow \expo{(x+m_i)^{-1}}{\psimorph(P_2)}$  & \textbf{return} $1$ \textbf{if} $(R,Q) \in \GG_1 \times \GG_2$ \\

      $(m,R,S,T)
      \xleftarrow{R} \mathcal{A}(\params,P_2,X,Y,m_1,\ldots,$ &  
      \phantom{\textbf{return} $1$ \textbf{if}}
      \textbf{and}  $\couple{R}{Q}=\couple{\psimorph(P_2)}{P_2}^{xyz}$ \\
      \phantom{$(m,R,S,T) \xleftarrow{R} \mathcal{A}(\params,$} $m_{\ell(k)},R_1,\ldots,R_{\ell(k)})$ & \phantom{\textbf{return}} $0$ \textbf{otherwise} \\

      \textbf{return} $1$ \textbf{if} $(R,S,T) \in \GG_1^3$   &  \\
      \phantom{\textbf{return} $1$ \textbf{if}} \textbf{and} satisfies (\ref{eq:DVSBBpreuve}) & \\

      \phantom{\textbf{return}} $0$ \textbf{otherwise} & \\
    \end{tabular}
  \end{center}

  \smallskip
  \noindent Let $\tau \in {\NN}^{\NN}$, $\eps \in [0,1]^{\NN}$ and let
  $i \in \{1,2\}$. We define the \emph{successes} of $\mathcal{A}$
  \emph{via} $$
  \mathbf{Succ}_{\textsf{Gen},\mathcal{A}}^{\probleme}(k) =
  \Pr[\mathbf{Exp}_{\textsf{Gen},\mathcal{A}}^{\probleme}(k)=1]. $$

  \begin{enumerate}
  \item $\Adver$ is a \emph{$\tau$-\probleme-adversary} if for all
    $k \in \mathbb{N}$, the experiment
    $\mathbf{Exp}_{\textsf{Gen},\mathcal{A}}^{\probleme}(k)$ ends in
    expected time less than $\tau(k)$.
  \item $\gen$ is a \emph{$(\tau,\eps)$-\probleme-secure-generator} if
    for any {$\tau$-\probleme-adversary} $\Adver$ and any $k \in \mathbb{N}$, $$\mathbf{Succ}_{\gen,\Adver}^{\probleme}(k) \leq
    \eps(k).$$
  \end{enumerate}
\end{definition}

\bigskip
\begin{definition} Let $\Adver$ be a PPTM. We consider the following
  random experiments, where $k \in \NN$ is a security parameter:

  \smallskip
  \begin{center} \begin{tabular}{l} \fbox{Experiment
        $\mathbf{Exp}_{\textsf{Gen},\mathcal{A}}^{\dsfflex\hbox{-}d}(k)$}~ \\

      $P_2 \xleftarrow{R} \GG_2 \setminus \{\mathbb{O}_{\GG_2}\}$  ; $(x,y,z,t) \xleftarrow{R} \range{q-1}^4$  \\

      $X \leftarrow \expo{x}{\psimorph(P_2)}, Y \leftarrow \expo{y}{P_2}, Z \leftarrow \expo{z}{P_2}$ \\ 

      \textbf{if} $d=0$ \textbf{then} $(R,Q) \leftarrow (\expo{xt}{\psimorph(P_2)},\expo{yzt^{-1}}{P_2})$ \textbf{otherwise} $(R,Q) \xleftarrow{R}
      \GGa \times \GGb$ \\

      $b \xleftarrow{R} \mathcal{A}(\bilinear,X,Y,Z,R,Q)$ \\

      \textbf{return} $1$ \textbf{if} $b=d$ \\ \phantom{\textbf{return}} $0$ \textbf{otherwise} \\

    \end{tabular}
  \end{center}

  \smallskip\noindent Let $\tau \in {\NN}^{\NN}$, $\eps \in
  [0,1]^{\NN}$. We define the \emph{advantage} of $\mathcal{A}$
  \emph{via}
$$
\mathbf{Adv}_{\textsf{Gen},\mathcal{A}}^{\dsfflex}(k) = \left\vert
  \Pr[\mathbf{Exp}_{\textsf{Gen},\mathcal{A}}^{\dsfflex-0}(k)=1] -
  \Pr[\mathbf{Exp}_{\textsf{Gen},\mathcal{A}}^{\dsfflex-1}(k)=1]
\right\vert. $$

\begin{enumerate}
\item $\Adver$ is a \emph{$\tau$-\dsfflex-adversary} if for all
  $k \in \mathbb{N}$, the experiment
  $\mathbf{Exp}_{\textsf{Gen},\mathcal{A}}^{\dsfflex}(k)$ ends in
  expected time less than $\tau(k)$.
\item $\gen$ is a \emph{$(\tau,\eps)$-\dsfflex-secure-generator} if
  for any {$\tau$-\dsfflex-adversary} $\Adver$ and any $k \in \mathbb{N}$, $$\mathbf{Adv}_{\gen,\Adver}^{\dsfflex}(k) \leq
  \eps(k).$$
\end{enumerate}
\end{definition}

\section{Description of the new schemes}
In this section, we describe our new UDVS schemes. The general
principle underlying the construction of \bbudvs{} and \blsudvs{} is
based on an elegant technique proposed by Damg\aa{}rd
\cite{Damgard1991} and aimed at making public-key encryption scheme
secure against (non-adaptive) chosen ciphertext attacks. We give in
details the ideas underlying their design, since we are convinced that
they may be of independent interest\footnote{Since the publication of
  \cite{Vergnaud2006a}, Laguillaumie, Libert and Quisquater
  \cite{LaguillaumieLibertQuisquater2006} have proposed new universal
  designated verifier signatures. The technique presented in this
  paper can be used to improve the efficiency of their schemes.}
(\eg{} for the construction of new privacy-preserving signature
schemes).

\subsection{Design principle}
Let $(\GG,+)$ be a group of prime order $q$ and let $P$ be a generator
of $G$. In 1991, Damg\aa{}rd \cite{Damgard1991} presented a simple
variant of the Elgamal encryption scheme in $\GG$. In his proposal,
Alice publishes two public keys $A_1 = \expo{a_1}{P}$ and $A_2 =
\expo{a_2}{P}$ and keeps secret their discrete logarithms $a_1$ and
$a_2$. When Bob wants to privately send a message $M \in \GG$ to
Alice, he picks uniformly at random an integer $r \in \range{q-1}$ and
transmits the triple $(Q_1,Q_2,C)$ where $Q_1 = \expo{r}{P}$, $Q_2 =
\expo{r}{A_1}$ and $C = M + \expo{r}{A_2}$. When she receives the
ciphertext $(Q_1, Q_2, C)$, Alice checks whether the equality $Q_2 =
\expo{a_1}{Q_1}$ holds: if it is the case, she retrieves the message
$M$, as $M = C - \expo{a_2}{Q_1}$, otherwise she rejects the
ciphertext.

Damg\aa{}rd proved that if the \DDH\ problem is hard in $\GG$, then
this scheme is semantically secure against (non-adaptive) chosen
ciphertext attacks, if we assume the so-called {knowledge-of-exponent
  assumption} \cite{BellarePalacio2004}.  Intuitively this assumption
states that, without the knowledge of $a_1$, the only way to generate
couples $(Q_1,Q_2) \in \GG^2$, verifying $Q_2 = \expo{a_1}{Q_1}$, is
to choose an integer $r \in \range{q-1}$ and to compute $Q_1 =
\expo{r}{P}$ and $Q_2 = \expo{r}{A_1}$.

There are many ways in which the formulation of KEA can be varied to
capture this intuition that the only way to generate a Diffie-Hellman
triple is to know the corresponding exponent
\cite{BellarePalacio2004,Damgard1991}.  Usually, this is done by
saying that for any PPTM outputting such a triple, there is an
``extractor'' than can return this exponent. For our purposes, it is necessary to allow the adversary to be
  randomized as in \cite{BarakLindellVadhan2003} (in that case, it is
  important that the extractor gets the coins $\varpi$ of the adversary as an
  additional input, since otherwise the assumption is clearly false).
We propose a similar
definition suitable for bilinear structures.

\begin{definition} Let $\bfA$ and $\bfAbar$ be two PPTM's. We consider
  the following random experiments, where $k \in \NN$ is a security
  parameter:

  \smallskip
  \begin{center} \begin{tabular}{l} \fbox{Experiment
      $\mathbf{Exp}_{\gen,\bfA,\bfAbar}^{\textsf{kea}}(k)$}~ \\
      $\bilinear \xleftarrow{R} \gen(k)$ \\

      $P_2 \xleftarrow{R} \GG_2
      \setminus \{\mathbb{O}_{\GG_2}\}$ ; $x \xleftarrow{R} \range{q-1}$ \\

      $(R, S) \xleftarrow{R} \bfA_k(\bilinear, P_2, \expo{x}{P_2})$ \\

      $r \leftarrow \bfAbar_k(\bilinear, P_2, \expo{x}{P_2};\varpi)$ \\

      \textbf{return} $1$ \textbf{if} $(R, S) \in \GG_1 \times \GG_2$, $\psimorph(S) =
      \expo{x}{R}$ \textbf{and} $R \neq \expo{r}{P_2}$ \\
      \phantom{\textbf{return}} $0$ \textbf{otherwise} \\
    \end{tabular} \end{center} We define the \emph{advantage} of $\bfA$
  relative to $\bfAbar$ \emph{via} $$
  \mathbf{Adv}_{\gen,\bfA,\bfAbar}^{\textsf{kea}}(k) =
  \Pr\left[\mathbf{Exp}_{\gen,\bfA,\bfAbar}^{\textsf{kea}}(k)=1\right].
$$

\noindent Let $\eps \in [0,1]^{\NN}$
\begin{enumerate}
\item $\bfAbar$ is a \emph{$\eps$-\textsf{kea}-extractor} for $\bfA$
  if for all $k \in \mathbb{N}$,
  $\mathbf{Adv}_{\gen,\bfA,\bfAbar}^{\textsf{kea}}(k) \leq \eps(k)$
\item We say that the knowledge-of-exponent assumption holds for
  $\gen$ if there exists a PPTM $\bfAbar$ such that for every PPTM
  $\bfA$, there exists a negligible function $\eps$ such that
  $\bfAbar$ is a \emph{$\eps$-\sfkea-extractor} for $\bfA$.
\end{enumerate}

\end{definition}

\subsection{Description of the protocol \bbudvs}\label{ssec:bbblind} \subsubsection{Boneh-Boyen's
  signatures.} In 2004, Boneh and Boyen \cite{BonehBoyen2004} proposed
a new application of bilinear structures to construct efficient short
signatures. Their idea is to plug the message to be signed in the
exponent and, in order to avoid trivial ``homomorphic'' forgeries, to
do so in a non-linear way. For an entity whose private/public key pair
is $(u, \expo{u}{P_2})$ in $\range{q-1} \times \GG_2$, the publication
of the group element $\sig = \expo{(u+m)^{-1}}{P_1}$ seems to be a
good mean to authenticate a message $m \in \range{q-1}$. Indeed, the
computation of $\sig$ for a given couple $(m, \expo{u}{P_2})$ is
equivalent to the resolution of the so called co-\CDH\ problem
\cite{BonehLynnShacham2004} and it seems to remain difficult even if
the adversary is allowed to choose $m$ and knows
$\expo{(u+m_1)^{-1}}{P_1}, \ldots, \expo{(u+m_s)^{-1}}{P_1}$ in $\GG$,
for some $m_i \in\range{q-1} \setminus \{m\}$ ($i \in \range{s}$).
Boneh and Boyen have proved that this problem is not easier than the \bbprob{(s+1)}
problem. They also made the important remark that the use of a second
pair of keys $(v, \expo{v}{P_2})$ in $\range{q-1} \times \GG$ enables
to prove the unforgeability of the scheme under chosen message
attacks, in the standard security model: they
suggested to replace the signature $\sig$ by a couple
$(\expo{(u+m+rv)^{-1}}{P_1}, r)$ where $r$ is picked uniformly at
random in $\range{q-1}$. Finally, in order to be able to sign
arbitrarily long messages, an hash function family $\hashfam$ is added
to the public parameters, such that for every group order $q$ output
by $\gen$, $\hashfam(q)$ generates the description of a (collision
resistant) hash function $\hash$ which maps arbitrary long bit strings
on elements from $\range{q-1}$.

\subsubsection{The scheme \bbudvs.} Let $\params = (\bilinear, P_1,
P_2)$ and $\hashfam$ be as above, let $(U_a, V_a) \in \GGb$ be Alice's
public key for \bobo{} signatures. The principle underlying the
universal designated verifier signature scheme \bbudvs{} is based on
Damg\aa{}rd's idea. Let us suppose that Bob has published a public key
$U_b = \expo{u_b}{P_2}$ and that the pair $\sig = (r, S)$ in
$\range{q-1} \times \GG_1$ is a \bobo\ signature produced by Alice, on
a message $m$. If Cindy wants to designate $\sig$ to Bob, she picks
uniformly at random an integer $t \in \range{q-1}$ and sets $Q_1 =
\expo{t}{S},$ $Q_2 = \expo{t}{U_b}$ and $Q_3 = \expo{t}{P_2}.$ The
quadruple $\dsig = (r, Q_1, Q_2, Q_3)$ is the resulting designated
verifier signature on $m$. The protocol \bbudvs{} is described with
all the details in figure \ref{fig:BBUDVS}. The following simple
observations are intuitive arguments in favor of the security of the
protocol.

\begin{figure}[t]
  \begin{center}
    \begin{tabular}{l}
      \hline
      \algo{} \bbudvs.\setup \\
      \hline
      \entree{} $k$ \\
      \sortie{} $\params$ \\
      \hline
      $\bilinear \hasard \textsf{Gen}(k)$ \\
      $\Pb \hasard \GGb$ ; $\Pa \leftarrow \psimorph(\Pb)$
      $\hash \hasard \hashfam(q-1)$ \\
      $\params \leftarrow [\bilinear,\Pa,\Pb,\hash]$ \\
      \hline
      \\
      \hline
      \algo{} \bbudvs.\sign \\
      \hline
      \entree{} \params, $m$, $(u_a,v_a)$ \\
      \sortie{} $\sig$ \\
      \hline
      $h \leftarrow \hash(m)$ \\
      \textsf{repeat} $r \hasard \range{q-1}$ \\
      \textsf{until} $u_a+h+v_ar \neq 0 \mod q$ \\
      $S \leftarrow \expo{(u_a+h+v_ar)^{-1}}{\Pa}$ \\
      $\sig \leftarrow (r, S)$ \\
      \hline
      \\
      \hline
      \algo{} \bbudvs.\vkeygen \\
      \hline
      \entree{} \params \\
      \sortie{} $(\skV,\pkV)$ \\
      \hline
      $u_b \hasard \range{q-1}$ \\
      $\skV \leftarrow u_b$ \\
      $\pkV \leftarrow \expo{u_b}{P_2}$ \\
      \hline
      \\
      \hline
      \algo{} \bbudvs.\fake \\
      \hline \entree{}
      \params, $m$, $u_b$, $(U_a,V_a)$ \\
      \sortie{} $\dsig$ \\
      \hline $(r,t) \hasard \range{q-1}^2$ \\
      $R \leftarrow \expo{t}{\left( \psimorph(U_a) + \expo{\hash(m)}{\Pa} + \expo{r}{\psimorph(V_a)}\right)}$ \\
      $Q_1 \leftarrow \expo t {P_1}$, $Q_2 \leftarrow \expo{u_b} R$,  $Q_3
      \leftarrow R$ \\
      $\dsig \leftarrow  (r,Q_1,Q_2,Q_3)$ \\
      \hline
    \end{tabular}
    \begin{tabular}{l}
      \hline
      \algo{} \bbudvs.\skeygen \\
      \hline
      \entree{} \params \\
      \sortie{} $(\skS,\pkS)$ \\
      \hline
      $(u_a,v_a) \hasard \range{q-1}^2$ \\
      $\skV \leftarrow (u_a,v_a)$ \\
      $\pkV \leftarrow (\expo{u_a}{P_2},\expo{v_a}{P_2})$ \\
      \hline
      \\
      \hline
      \algo{} \bbudvs.\verify \\
      \hline \hline
      \entree{} \params, $m$, $(U_a,V_a)$, $(r,S)$ \\
      \sortie{} $b$  \\
      \hline
      $\alpha \leftarrow \couple{S}{U+\expo{\hash(m)}{\Pb}+\expo{r}{V}}$ \\
      \sialorssinon{$\alpha = \couple{\Pa}{\Pb}$}{$b \leftarrow \valid$}{$b \leftarrow
        \invalid$} \\
      \hline
      \\
      \hline
      \algo{} \bbudvs.\designate \\
      \hline
      \entree{} \params, $m$, $(U_a,V_a)$,  $U_b$, $(r,S)$ \\
      \sortie{} $\dsig$ \\
      \hline
      $t \hasard\range{q-1}$ \\
      $Q_1 \leftarrow \expo{t}{S}$, \ $Q_2 \leftarrow
      \expo{t}{\psimorph(U_b)}$, \\ $Q_3 \leftarrow \expo{t}{P_1}$
      \\
      $\dsig \leftarrow  (r,Q_1,Q_2,Q_3)$ \\
      \hline

      \\

      \hline \algo{} \bbudvs.\dverify \\ \hline \entree{}
      \params, $m$, $u_b$, $(U_a,V_a)$, \\
      \phantom{\entree}
      $(r,Q_1,Q_2,Q_3)$ \\ \sortie{} $b$
      \\ \hline $\alpha_1 \leftarrow
      \couple{Q_1}{U_a+\expo{\hash(m)}{\Pb}+\expo{r}{V_a}}$ \\ $\alpha_2
      \leftarrow \couple{Q_3}{\Pb}$ \\ $\beta_1 \leftarrow
      \couple{Q_3}{\expo{u_b}{P_2}}$,  $\beta_2 \leftarrow
      \couple{Q_2}{P_2}$
      \\ \sialorssinonl{$\alpha_1 = \alpha_2 \land \beta_1 = \beta_2$}{$b
        \leftarrow \valid$}{$b \leftarrow \invalid$} \\
      \hline
    \end{tabular}
  \end{center}
  \caption{\descprot{} $\bbudvs(\gen,\hashfam)$ \label{fig:BBUDVS}}
\end{figure}

\begin{enumerate}
\item Under KEA, the equality
  \begin{equation}\label{eq:eq1}
    \couple{Q_3}{U_b} = \couple{Q_2}{\Pb}
  \end{equation}
  insures Bob that Cindy knows the value $t$ such that $Q_2 =
  \expo{t}{U_b}$ and $Q_3 = \expo{t}{P_2}$.
\item If (\ref{eq:eq1}) is satisfied, Bob is convinced that Cindy
  knows the group element $S = \expo{t^{-1}}{Q_1}$. The \bobo\
  verification equality
  $\couple{S}{U_a+\expo{\hash(m)}{\Pb}+\expo{r}{V_a}} =
  \couple{\Pa}{\Pb}$, holds if and only if the equality
  \begin{equation}\label{eq:eq2}
    \couple{Q_1}{U_a+\expo{\hash(m)}{\Pb}+\expo{r}{V_a}} =
    \couple{Q_3}{\Pb}
  \end{equation}
  does. Therefore, if the equalities (\ref{eq:eq1}) and (\ref{eq:eq2})
  are true, the quadruple $\dsig$ proves to Bob that Alice has
  actually signed the message $m$.
\item However, this quadruple cannot convince anyone else, since it
  could have been produced by Bob himself. Indeed, if Bob samples
  uniformly at random $(r,\tilde t)$ in $\range{q-1}^2$ and computes
  the group elements:
$$Q_1 =  \expo{\tilde{t}}{P_1},
Q_2 = \expo{u_b \tilde{t}}{\left({U_a} + \expo{\hash(m)}{\Pb} +
    \expo{r}{V_a} \right)}, Q_3 = \expo{\tilde{t}}{U_a} +
\expo{\tilde{t} \cdot \hash(m)}{\Pb} + \expo{\tilde{t} \cdot
  r}{V_a},$$ he produces quadruples which verify (\ref{eq:eq1}) and
(\ref{eq:eq2}) and follow the same distribution as those produced by
Cindy (namely with $t \equiv_q \tilde{t} (u_a + \hash(m) + v_a r)$).
\end{enumerate}

\remark{Given a UDVS produced by \bbudvs{}, it is easy, by random
  scalar multiplication, to produce a new signature on \emph{the same}
  message for the \emph{same} public keys. It is admitted that weak
  forgery is no real threat whatsoever.}

\bigskip
\remark{The computational workload of $\bbudvs.\dverify$ for the
  designated verifier can be reduced to only two pairing evaluations
  and one bilinear exponentiation thanks to the knowledge of $u_b$ by
  checking that $Q_2 = \expo{u_b}{Q_3}$ instead of $\beta_1 =
  \beta_2$.}

\bigskip
\remark{In the algorithm $\bbudvs.\fake$, the verifier's secret key
  $u_b$ is used only to compute $Q_2 = \expo{u_b}{R}$. Therefore, the
  signer as well as the verifier can delegate his authenticating
  capacity (without revealing the secret key) by publishing the
  elements $K_1 = \expo{u_a \cdot u_b}{\Pa}$ and $K_2 = \expo{v_a
    \cdot u_b}{\Pa}$ in $\GGb$. Indeed, the knowledge of $(K_1,K_2)$
  suffices to produce an UDVS $\dsig = (r, Q_1,Q_2, Q_3)$ on a message
  $m$ by picking uniformly at random $(r,t) \in \range{q-1}^2$, and
  computing $Q_1 \leftarrow \expo t {P_1}$, $Q_2 \leftarrow
  \expo{t}{K_1} + \expo{t \cdot \hash(m)}{\psimorph(U_b)} + \expo{t
    \cdot r}{K_2}$ and $Q_3 = \expo{t}{\psimorph(U_a)} + \expo{t \cdot
    \hash(m)}{\Pa} + \expo{t \cdot r}{\psimorph(V_a)}$. Therefore, the
  scheme $\bbudvs$ is delegatable.}

\subsection{Description of the protocol
\blsudvs}\label{ssec:blsblind} \subsubsection{Boneh-Lynn-Shacham's
signatures.} In \cite{BonehLynnShacham2004}, Boneh \etali{} presented
the signature scheme \bls\ that works in any bilinear cryptographic
context. The scheme resembles the undeniable signature scheme proposed
by Chaum and van Antwerpen \cite{ChaumvanAntwerpen1990} and can be
seen as a variant of the FDH signature scheme
\cite{BellareRogaway1993}. The protocol \bls{} is efficient, produces
short signatures (for carefully chosen parameters), and is unforgeable
in the random oracle model assuming the intractability of the co-\CDH\ problem.

\subsubsection{The scheme \blsudvs.} Let \gen\ be a
prime-order-BDH-para\-meter-generator, let $f_r \in {\NN}^{\NN}$, and
let $\HashFam$ be an hash function family such that for bilinear
structure $\bilinear$ output by $\gen$, $\HashFam(\GGa)$ generates the
description of an hash function $\Hash$ (modeled in the security
analysis as a random oracle) which maps arbitrary long bit strings on
elements from $\GGa$. Let \bls\ be the associated signature scheme;
using the same approach, it is possible to construct a new UDVS scheme
compatible with the $\bls$ signatures. The protocol \blsudvs{} is
described with all the details in figure \ref{fig:BLSUDVS}.

\begin{small}
  \begin{figure}[t]
    \begin{center}
      \begin{tabular}{l}
        \hline
        \algo{} \blsudvs.\setup \\
        \hline
        \entree{} $k$ \\
        \sortie{} $\params$ \\
        \hline
        $\bilinear \hasard \textsf{Gen}(k)$ \\
        $\Pb \hasard \GGb$ ;
        $n_r \leftarrow f_r(k)$ ;
        $\Hash \hasard \HashFam(\GGa)$ \\
        $\params \leftarrow [\bilinear,\Pb,n_r,\Hash]$ \\
        \hline
        \\
        \hline
        \algo{} \blsudvs.\sign \\
        \hline
        \entree{} \params, $m$, $u$ \\
        \sortie{} $\sig$ \\
        \hline
        $r \hasard \{0,1\}^{n_r}$ ;
        $H \leftarrow \Hash(m,r)$ \\
        $S \leftarrow \expo{u}{H}$ ;
        $\sig \leftarrow  (r,S)$ \\
        \hline
        \\
        \hline
        \algo{} \blsudvs.\vkeygen \\
        \hline
        \entree{} \params \\
        \sortie{} $(\skV,\pkV)$ \\
        \hline
        $\skV = u_b \hasard \range{q-1}$ \\
        $\pkV = U_b \leftarrow \expo{u_b}{\Pb}$ \\
        \hline
        \\
        \hline
        \algo{} \blsudvs.\fake \\
        \hline
        \entree{} \params, $m$, $\pkS$, $\skV$ \\
        \sortie{} $\dsig$ \\
        \hline $r \hasard \{0,1\}^{n_r}$ ;
        $t \hasard \range{q-1}$ \\
        $Q_1 \leftarrow \expo{t^{-1}}{H(m,r)}$ ;
        $Q_2 \leftarrow \expo{t \cdot \skV}{\pkS}$ \\
        $\dsig \leftarrow  (r,Q_1,Q_2)$ \\
        \hline
      \end{tabular}
      \begin{tabular}{l}
        \hline
        \algo{} \blsudvs.\skeygen \\
        \entree{} \params \\
        \sortie{} $(\skS,\pkS)$ \\
        \hline
        $\skS = u_a \hasard \range{q-1}$ \\
        $\pkS = U_a \leftarrow \expo{u_a}{\Pb}$ \\
        \hline
        \\
        \hline
        \algo{} \blsudvs.\verify \\
        \hline
        \entree{} \params, $m$, $U_a$, $(r,S)$ \\
        \sortie{} $b$  \\
        \hline
        $H \leftarrow \Hash(m,r)$ \\
        $s \leftarrow \couple{H}{U_a}$ \\
        \sialorssinon{$s = \couple{S}{\Pb}$}{$b \leftarrow \valid$}{$b
          \leftarrow \invalid$} \\
        \hline
        \\
        \hline
        \algo{} \blsudvs.\designate \\
        \hline
        \entree{} \params, $m$, $\pkS$, $(r,S)$, $\pkV$ \\
        \sortie{} $\dsig$ \\
        \hline
        $t \hasard \range{q-1}$ \\
        $Q_1 \leftarrow \expo{t}{S}$ ;
        $Q_2 \leftarrow \expo{t^{-1}}{\pkV}$ \\
        $\dsig \leftarrow  (r,Q_1,Q_2)$ \\
        \hline
        \\
        \hline
        \algo{} \blsudvs.\dverify \\
        \hline
        \entree{} \params, $m$, $\pkS$, $(r, Q_1, Q_2)$,  $\skV$ \\
        \sortie{} $b$  \\
        \hline $H \leftarrow H(m,r)$
        $s \leftarrow \couple{\expo{\skV}{H}}{\pkS}$ \\
        \sialorssinonl{$s = \couple{Q_1}{Q_2}$}{$b \leftarrow \valid$}{$b
          \leftarrow \invalid$} \\
        \hline
      \end{tabular}
    \end{center}
\caption{\descprot{} $\blsudvs(\gen, f_r, \HashFam)$
  \label{fig:BLSUDVS}}
\end{figure}
\end{small}

\noindent Let $k \in \NN$, let $\params = (\bilinear, P_2, \Hash)$ be
some output of $\bls. \setup(k)$ and let $U_a = \expo{u_a}{P_2}$
(\resp\ $U_b = \expo{u_b}{P_2}$) be Alice's (\resp\ Bob's) public key
output by $\bls.\keygen(\params)$. \bls\ signatures are elements $S =
\expo{u_a}{H} \in \GG_1$, where the group element $H$ is the hash
value of the signed message $m$ and (potentially) some random salt (of
size $n_r=f_r(k)$). The discrete logarithm of $H$ is unknown to all
users, therefore, whence the signature $S$ is randomized as above:
$Q_1 = \expo{t}{S}$ for some $t \in \range{q-1}$, it suffices to
reveal the element $Q_2 = \expo{t^{-1}}{U_b}$ to prove, in a
non-transferable way, to Bob that Alice actually signed the message
$m$. The tuple $(P_2, U_a, U_b, H, \couple{Q_1}{Q_2})$ is indeed a
bilinear Diffie-Hellman tuple which could have been produced by using
secret information from Alice or Bob, but not otherwise under the
assumption that the computational bilinear Diffie-Hellman problem
problem is intractable.

\bigskip \remark{The protocol $\blsudvs$ is delegatable
  \cite{LipmaaWangBao2005}.  Indeed, in the algorithm
  $\blsudvs.\fake$, the secret key $u_b$ from the designated verifier
  is only used to compute the element $Q_2 = \expo{t \cdot u_b}{U_a}
  \in \GGb$ and the signer as well as the verifier can delegate their
  authenticating capability (without disclosing their secret key) by
  publishing the element $\expo{u_a \cdot u_b}{\Pb}$.}

\section{Security results} In this section, we state the security
properties of our schemes.
\subsection{Unforgeability of the scheme \bbudvs{}}
The theorem below states that the protocol $\bbudvs(\gen, \hashfam)$
is \UDVSEFCMA-secure assuming the \COKEA{} assumption, the
collision-resistance of $\hashfam$ and the intractability of the
problem $\bbprob{\ell}$ in $\gen$, for all polynomial $\ell \in
{\NN}^{\NN}$. Since KEA is a somewhat strange and impractical
assumption, it would be better if we could do without it, as it has
been recently done by Gj\o{}steen \cite{Gjosteen2006} for
Damg\aa{}rd's encryption scheme. In the following theorem, we prove
the unforgeability of \bbudvs{} to \dvsbbprobcourt{\ell} without KEA.
Finally, since the protocol \bbudvs{} is publicly verifiable, we
consider only \UDVSEFCMA-attackers that do not make queries to the
verifying oracle \orCon.

\bigskip
\begin{theorem}\label{the:bbudvs}%
  \newcommand{\schema}{\bbudvs}
  Let \gen{} be a prime-order-BDH-generator and $\hashfam$ be an
  hash-function family of codomain indexed by the orders of groups
  generated by $\gen$.
  \begin{enumerate}%
  \item If the scheme $\bobo(\gen, \hashfam)$ is \EFCMA-secure against
    polynomial adversaries, then under the \COKEA{} assumption in
    \gen, the scheme $\schema(\gen, \hashfam)$ is \UDVSEFCMA-secure
    against polynomial adversaries.
  \item If for all polynomial $\ell$, $\gen$ is $\bbprob{\ell}$-secure
    against polynomial adversaries and if $\hashfam$ is an
    hash-function collision-resistant against polynomial adversaries
    then, under the \COKEA{} assumption in \gen, the protocol
    $\schema(\gen, \HashFam)$ is \UDVSEFCMA-secure against polynomial
    adversaries.
    \newcommand{\probl}{\dvsbbprobcourt{\qSig}}%
  \item Let $(\tau,\qSig) \in \fonction{\NN}{\NN}^2$ and let $\Adver$ be a
    $(\tau,\qSig,0)$-\UDVSEFCMA-adversary against the scheme $\schema(\gen,
    \hashfam)$.  There exist $\tau', \tau'' \in \fonction{\NN}{\NN}$
    verifying,
$$
\tau' = \tau + \qSig \cdot (\Texp{\GGa} + O(1)) \hbox{ and } \tau'' =
\tau + O(1),
$$
a $\tau'$-\probl-adversary $\AdverB$ against \gen{} and a
$\tau''$-\COLL-adversary $\AdverC$ against $\hashfam$ such that,
$$2 \cdot \Succ{\gen}{\AdverB}{\probl} + \Succ{\hashfam}{\AdverC}{\COLL}
\geq \Succ{\schema}{\Adver}{\UDVSEFCMA}.$$
\end{enumerate}
\end{theorem}

\bigskip \textit{Proof.}
{\begin{enumerate}
  \item The algorithm $\AdverB$ which try to forge a signature
    $\bobo$, takes as input some public parameters $\params$ and a
    signing public key $\pks$. It computes a verifying public key $U_b
    = \expo{u_b}{\Pb}$ by running the algorithm
    $\bbudvs.\vkeygen'(\params)$ and then executes the algorithm
    $\Adver$ on the entries $\params$, $\pks$ and $U_b$. It forwards
    $\Adver$'s signature queries to its own signing oracle and the
    simulation of the verifying oracles is straightforward since the
    protocol $\bbudvs$ is publicly verifiable.

    \medskip Let us denote $\Adver'$ the algorithm whose execution is
    identical to the one of $\Adver$, but which returns the pair
    $(Q_3, Q_2)$, when $\Adver$ returns $\dsigst = (r, Q_1, Q_2,
    Q_3)$.  If $\Adver$'s output is a valid forgery, then the
    $4$-tuple $(\psimorph(P_2), \expo{u_b}{\psimorph(P_2)}, Q_3, Q_2)$ is a valid
    Diffie-Hellman $4$-tuple.  Assuming \COKEA{}, there exists
    $\overline{\Adver'}$ which taken as inputs $\Adver'$'s random tape
    and entries, outputs $t \in \range{q-1}$ such that $Q_3 =
    \expo{t}{P_2}$ et $Q_2 = \expo{t}{U_b}$ with a probability
    negligibly close to the success of $\Adver$.

    \medskip $\AdverB$ run the algorithm $\overline{\Adver'}$ to get
    this value $t$ and outputs the pair $\sigst = (r,
    \expo{t^{-1}}{Q_1})$ which is a valid forgery for the scheme
    $\bobo$ if $\dsig$ is a valid forgery and $Q_3 = \expo{t}{P_2}$.
    The probability of success of $\AdverB$ is therefore negligibly
    close to the one of $\Adver$ and its running time is polynomial.

  \item It is a simple consequence of the first part of the theorem
    and the security theorem from \cite{BonehBoyen2004}.

  \item
    \newcommand{\schema}{\bbudvs}
    \newcommand{\probl}{\dvsbbprobcourt{\qSig}}
    Let $\gen$ be a prime-order-BDH-generator, $\qSig,\tau \in
    \fonction{\NN}{\NN}$ and let $\Adver$ be a
    $(\tau,\qSig,0)$-\UDVSEFCMA-adversary against $\schema(\gen)$. It
    is readily seen that $\Adver$ can be converted into an
attacker for the simplified scheme defined without the hash function
$\hashfam$ or into an attacker $\mathcal{C}$ against the collision
resistance of $\hashfam$. For the sake of simplicity, we will assume that
the scheme works directly on messages $m \in \range{q}$. 

    \medskip We will construct an algorithm $\AdverB$ which takes as
    inputs $\bilinear$ generated by $\gen(k)$, a vector $(m_1, \ldots,
    m_{\qSig(k)}) \in \range{q}^{\qSig(k)}$, $(\Pb,X,Y) \in \GGb^3$
    and $(R_1,\ldots,R_{\qSig(k)}) \in \GGa^{\qSig(k)}$ satisfying
    $R_i = \expo{(x+m_i)^{-1}}{\Pa}$ for all $i \in \range{\qSig(k)}$
    with $\Pa = \psimorph(\Pb)$ et $X = \expo{x}{\Pb}$, outputs a
    $4$-tuple
$$(m,R,S,T) \in \left(\range{q-1} \setminus
  \{m_1,\ldots,m_{\qSig(k)}\}\right) \times \GGa^3$$ which satisfies
(\ref{eq:DVSBBpreuve}).

Our method of proof is inspired by Shoup \cite{Shoup2002}: we define a
sequence of games $\Exp_1$, \ldots, $\Exp_{4}$ starting from the
actual \UDVSEFCMA-adversary $\Adver$ and modify it step by step, until
we reach a final game whose success probability has an upper bound
related to solving the \dvsbbprobcourt{\qSig} problem. All the games
operate on the same underlying probability space: the public and
private keys of the signature scheme and the coin tosses of ${\mathcal
  A}$.

\begin{Jeux} \item $\AdverB_1$ plays the role of the challenger in the
  experiment \Exper{\schema}{\Adver}{\UDVSEFCMA} of the definition
  \ref{def:UDVS-EFCMA}:

  \smallskip
  \begin{small} \begin{list}{}{} \item[\textsf{Initialization}] $k$ \\
      $\params \hasard \schema.\setup(k)$, \\ $(\sks,\pks) \hasard
      \schema.\skeygen(\params)$, $(\skc,\pkc) \hasard
      \schema.\vkeygen(\params)$ \\ \textbf{run}
      $\Adver(\params,\pks,\pkc)$ $\leadsto (\mst, \dsigst)$.

    \item [\textsf{Simulation of the oracles}] \quad \\
      \itemain{} $\orSig(m)$ $\leftrightsquigarrow$
      $\schema.\sign(\params,m,\sks)$. \\
    \end{list}
  \end{small}

  In the random experiments $\Exp_i$, for $i \in \range{5}$, we denote
  by $\mathsf{F_i}$, the event ``$\mst \notin \mathcal{Q}_{\orSig}$
  and $\schema.\dverify(\params,\mst,\pks,\dsigst,\pkc,
  \skc)=\valid$''. 

By definition, we have $\Pr[\mathsf{F_1}] =
  \Succ{\schema}{\Adver}{\UDVSEFCMA}(k).$

\item $\AdverB_2$ modify the previous simulation by inserting the
  bilinear structure underlying the instance of the problem \probl{}
  to solve in the public parameters $\params$.

  \smallskip
  \begin{small}
    \begin{list}{}{}%
    \item[\textsf{Initialization}] $k$ \\
      $\params \leftarrow [\bilinear, \Pa, \Pb]$, \\
      $(u_a, v_a, u_b) \hasard \range{q-1}^3$,
      $(U_a, V_a, U_b) \leftarrow (\expo{u_a}{\Pb},\expo{v_a}{\Pb},\expo{u_b}{\Pb})$ \\
      \textbf{run} $\Adver(\params,(U_a,V_a),U_b)$ $\leadsto (\mst,
      \dsigst)$.
    \item [\textsf{Simulation of the oracles}] \quad \\
      \itemain{} $\orSig(m)$ $\leftrightsquigarrow$ $\schema.\sign(\params,m,(u_a,v_a))$. \\
    \end{list}
  \end{small}

  The distribution of $\Adver$'s entries is unchanged and we have
  $\Pr[\mathsf{F_2}] = \Pr[\mathsf{F_1}].$

\item The algorithm $\AdverB_3$ precomputes the signatures given to
  the adversary $\Adver$ and then uses his knowledge of the secret key
  $u_a$ or $v_a$ in the chameleon hash function to answer $\Adver$'s
  signature queries. The algorithm $\AdverB_3$ distinguishes two types
  of forgers:

  \medskip
  \begin{itemize}
  \item[\textbf{Type 0:}] the forgers which
    \begin{enumerate}
    \item either make a signature query on $m$ such that $m=-u_a$;
    \item or return a forgery $(\mst,\dsigst)$ with $\dsigst =
      (r^{\star}, Q_1^{\star}, Q_2^{\star}, Q_3^{\star})$, such that
      $\mst + v_a r^{\star} \notin \{m_1, \ldots,m_{\qSig(k)} \}.$
    \end{enumerate}
  \item[\textbf{Type 1:}] the other forgers, namely those which
    \begin{enumerate}
    \item do not make a signature query on $m$ such that $m=-u_a$;
    \item and return a forgery $(\mst,\dsigst)$ with $\dsigst =
      (r^{\star}, Q_1^{\star}, Q_2^{\star}, Q_3^{\star})$, such that
      $\mst + v_a r^{\star} = m_i \hbox{ for some } i \in
      \range{\qSig(k)}.$
    \end{enumerate}
  \end{itemize}

  The adversary $\Adver$ is necessarily of one of these two types and
  the algorithm $\AdverB_4$ picks uniformly at random a bit $\beta \in
  \{0,1\}$. This algorithm will be able (at the end of the simulation)
  to solve the \probl{} problem if the adversary $\Adver$ is of type
  $\beta$.

  \smallskip
  \begin{small}
    \begin{list}{}{}%
    \item[\textsf{Initialization}] $k$ \\
      $\params \leftarrow [\bilinear, \Pa, \Pb]$, $c \leftarrow 1$,
      $\ell \leftarrow 0$, $\beta \hasard \{0,1\}$ \\
      $(u_a, v_a, u_b) \hasard \range{q-1}^3$ ;
      $(U_a, V_a, U_b) \leftarrow (\expo{u_a}{\Pb},\expo{v_a}{\Pb},\expo{u_b}{\Pb})$ \\
      $(h_1,\ldots,h_{\qSig(k)}) \hasard \range{q}^{\qSig(k)}$ \\
      \textbf{for} $i$ \textbf{from} $1$ \textbf{to} $\qSig(k)$ \textbf{do} \\
      \phantom{\textbf{for}} \textbf{if} $\beta=0$ \textbf{then} $T_i
      \leftarrow \expo{(u_a+h_i)^{-1}}{\Pa}$  \textbf{else} $T_i \leftarrow \expo{(v_a+h_i)^{-1}}{\Pa}$ \\
      \textbf{run} $\Adver(\params,(U_a,V_a),U_b)$ $\leadsto (\mst,
      \dsigst)$.
    \item [\textsf{Simulation of the oracles}] \quad \\
      \itemain{} $\orSig(m)$: \textbf{if} $\beta=0$ \textbf{and} $\expo{m}{\Pb} =  - U_a$ \textbf{then} $\ell \leftarrow - m \mod q$ \\
      \phantom{\itemain{} $\orSig(m)$:}
      \textbf{if} $\beta=0$ \textbf{then} $r \leftarrow (h_c-m) \cdot v_a^{-1} \mod q$, $S \leftarrow T_c$ \\
      \phantom{\itemain{} $\orSig(m)$: \textbf{if} $\beta=0$} \textbf{else} $r \leftarrow (u_a+m)\cdot h_c^{-1} \mod q$, $S \leftarrow \expo{r^{-1}}{T_c}$ \\
      \phantom{\itemain{} $\orSig(m)$:} \textbf{if} $r = 0$ \textbf{then} \textbf{return} $\perp$ \textbf{else} $c \leftarrow c+1$, \textbf{return} $(r,S)$. \\
    \end{list}
  \end{small}

  In both cases, the signatures produced by $\AdverB_3$ are perfectly
  distributed. We have indeed $\couple{S}{U_a + \expo{m}{\Pb} +
    \expo{r}{V_a}} = \couple{T_c}{U_a + \expo{h_c}{\Pb}} =
  \couple{\Pa}{\Pb}$, for $\beta=0$ and $\couple{S}{U_a +
    \expo{m}{\Pb} + \expo{r}{V_a}} =
  \couple{\expo{r^{-1}}{T_c}}{\expo{r \cdot h_c}{\Pb} + \expo{r}{V_a}}
  = \couple{\Pa}{\Pb}$ for $\beta=1$.

  In the random experiment $\Exp_i$, let us denote for $i \in \{ 3,4
  \}$, $\mathsf{T_i}$ the event ``$\Adver$ is of type $\beta$''. The
  algorithm $\AdverB_3$ aborts the simulation only if $\Adver$ is of
  type $1-\beta$. 

Therefore, we have $ \Pr[\mathsf{F_3} \vert
  \mathsf{T_3}] = {\Pr[\mathsf{F_2}]}.  $

\item $\AdverB_4$ replace in the following the public keys given as
  inputs to $\Adver$ and the precomputed signatures $(h_i,S_i)$ by
  elements coming from the instance of the problem to solve.

  \begin{small}
    \begin{list}{}{}
    \item[\textsf{Initialization}] $k$ \\%
      $\params \leftarrow [\bilinear, \Pa, \Pb]$, $c \leftarrow 1$,
      $\ell \leftarrow 0$, $\beta \hasard \{0,1\}$ \\\textbf{if} $\beta=0$ \textbf{then} $(U_a, U_b) \leftarrow (X,Y)$, $v_a \hasard \range{q-1}$, $V_a \leftarrow \expo{u_b}{\Pb}$  \\
      \phantom{\textbf{if} $\beta=0$} \textbf{else} $(V_a, U_b)
      \leftarrow (X,Y)$, $u_a \hasard \range{q-1}$, $U_a \leftarrow
      \expo{u_b}{\Pb}$ \\
      $(h_1,\ldots,h_{\qSig(k)}) \leftarrow (m_1,\ldots,m_{\qSig(k)})$ \\
      $(S_1,\ldots,S_{\qSig(k)}) \leftarrow (R_1,\ldots,R_{\qSig(k)})$ \\
      \textbf{run} $\Adver(\params,(U_a,V_a),U_b)$ $\leadsto (\mst,
      \dsigst)$.
    \end{list}
  \end{small}

  In the random experiment $\Exp_3$, if $\beta=0$ (\resp{} if
  $\beta=1$) the knowledge of $(u_a,u_b,v_b)$ (\resp{} of
  $(v_a,u_b,v_b)$) is not necessary to answer \Adver's signature
  queries. Hence, $\AdverB_4$ can still answer $\Adver$'s queries and
  since the distribution of the public keys and the precomputed
  signatures is unchanged, we get
$$\Pr[\mathsf{F_4} \vert \mathsf{T_4}]=\Pr[\mathsf{F_3} \vert
\mathsf{T_3}].$$

Eventually, when $\Adver$ returns the pair $(\mst, \dsigst)$, with
$\dsigst = (r^{\star}, Q_1^{\star}, Q_2^{\star}, Q_3^{\star})$, the
algorithm $\AdverB$ can solve the instance of the problem
\dvsbbprobcourt{\qSig(k)}:
\begin{itemize}
\item if $\Adver$ is of type $1$ and returns a forgery on a message
  $\mst$ satisfying the relation ${(m_i-\mst)\cdot (r^{\star})^{-1}} =
  x \mod q$ or if $\Adver$ is of type $0$ and has made a signature
  query on a message $m$ such that $m=-x$, then $\AdverB_4$ can
  retrieve the discrete logarithm of $X$ in base $\Pb$ and it can
  trivially produce a triple $(R,S,T)$ verifying the equality
  (\ref{eq:DVSBBpreuve});
\item otherwise, $\AdverB$ computes $m = \mst + r^{\star} v_a \mod q$
  and stops its execution by outputting the triple $(m,R,S,T) \in
  \range{q-1} \times \GGb \times \GGa \times \GGc$ where $m = \mst +
  r^{\star} v_a \mod q$ $R = Q_3^{\star}$, $S = Q_1^{\star}$ and $T =
  Q_2^{\star}$.
\end{itemize}

\begin{small}
  \begin{list}{}{}
  \item[\textbf{End of $\AdverB = \AdverB_4$'s execution}]
    $(\mst,\dsigst)$ \\%
    $(r^{\star}, Q_1^{\star}, Q_2^{\star}, Q_3^{\star}) \leftarrow \dsigst$ \\
    \textbf{if} $\beta=1$ \textbf{and} $\exists m_i, \expo{(m_i -
      \mst)\cdot (r^{\star})^{-1}}{\Pb}= U_a$ \textbf{do}
    $\ell \leftarrow (m_i-\mst)\cdot (r^{\star})^{-1}$ \\
    \textbf{if} $\ell \neq 0$ \textbf{then} $m \hasard \range{q-1} \setminus \{m_1,\ldots,m_{\qSig(k)}\}$ ;  $r \hasard \range{q-1}$ \\
    \phantom{\textbf{if} $\ell \neq 0$ \textbf{then}} $R \leftarrow
    \expo{r}{\Pa}$ ; $S \leftarrow \expo{r(\ell+m)^{-1}}{\Pa}$ ;
    $T \leftarrow \expo{r}{\psimorph(Y)}$ \\
    \textbf{if} $\beta=\ell=0$ \textbf{then} $m  \leftarrow  \mst + r^{\star} v_a \mod q$ \\
    \phantom{\textbf{if} $\beta=\ell=0$ \textbf{then}} $R  \leftarrow  Q_3^{\star}$ ; $S  \leftarrow  Q_1^{\star}$ ; $T  \leftarrow  Q_2^{\star}$ \\
    \textbf{return} $(m,R,S,T)$.
  \end{list}
\end{small}
\end{Jeux}

Clearly, if the event $\mathsf{F_4} \cap \mathsf{T_4}$ occurs, the
algorithm $\AdverB=\AdverB_4$ returns a $4$-tuple $(m,R,S,T)$ which
satisfies (\ref{eq:DVSBBpreuve}) and since $\Pr[\mathsf{T_4}] = 1/2$,
we get
$$
2 \cdot \Succ{\gen}{\AdverB}{\probl} + \Succ{\hashfam}{\AdverC}{\COLL}
\geq \Succ{\schema}{\Adver}{\UDVSEFCMA}.
$$
The algorithm $\AdverB$ runs in time less than $\tau(k) + \qSig(k)
(\Texp{\GGa} + O(1)),$ which concludes the proof.
\end{enumerate}
}

\subsection{Unforgeability and anonymity of the scheme \blsudvs} 
We prove (in the random oracle model) that \blsudvs{} is
\UDVSEFCMA-secure under the assumption that the problem \flex{} is
intractable in $\gen$. It is worth noting that this problem is at
least as hard as the computational bilinear Diffie-Hellman problem
underlying the schemes from
\cite{LaguillaumieVergnaud2005a,SteinfeldBullWangPieprzyk2003}.  We
prove also (again in the random oracle model) that the protocol
\blsudvs{} is \USPSICMA-secure under the assumption that the
decisional variant of this problem is intractable in $\gen$.

\bigskip
\begin{theorem}\label{the:BLSUDVS} %
  \newcommand{\schema}{\blsudvs}%
  Let $f_r \in \fonction{\NN}{\NN}$ and let \gen{} be a
  prime-order-BDH-generator. 

\noindent Let $(\qSig,\qCon,\qHash,\tau) \in
  \fonction{\NN}{\NN}^4$.
  \begin{enumerate} %
    \newcommand{\probl}{\flex}
  \item For all $(\tau,\qSig,\qCon)$-\UDVSEFCMA-adversary $\Adver$
    against $\schema(\gen, f_r, \orHash)$ where \orHash{} is a
    \qHash-random oracle, there exists $\tau' \in \fonction{\NN}{\NN}$
    verifying
$$ \tau' \leq \tau + (\qHash + \qSig + 2 \qCon + 2)(\Texp{\GGa} + O(1)) $$
and a $\tau'$-\probl-adversary $\AdverB$ against \gen{} such that
$$
\Succ{\gen}{\AdverB}{\probl} \geq
\frac{\Succ{\schema}{\Adver}{\UDVSEFCMA}}{(1 + 6 \cdot \qSig \cdot
  2^{f_r}) (\qCon+1)}. $$ %
\renewcommand{\probl}{\dsfflex}
\item For all $(\tau,\qSig,\qCon)$-\USPSICMA-distinguisher $\Adver$
  against $\schema(\gen, f_r, \orHash)$ where \orHash{} is a
  \qHash-random oracle, there exists $\tau' \in \fonction{\NN}{\NN}$
  verifying,
$$ \tau' \leq \tau + (\qHash + \qSig + 2 \qCon +
1)(\Texp{\GGa} + O(1)),$$ and a $\tau'$-\probl-distinguisher $\AdverB$
against \gen{} such that
$$ \Adv{\gen}{\AdverB}{\probl}
\geq \displaystyle{\frac{\Adv{\schema}{\Adver}{\USPSICMA}}{2} -
  \frac{\qSig+\qCon}{2^{f_r}}}. $$%
\end{enumerate}%
\end{theorem}

\bigskip \textit{Proof.}
{ \newcommand{\schema}{\blsudvs}%
  \newcommand{\probl}{\flex}%
  \begin{enumerate}
  \item

    \medskip The algorithm $\AdverB$, which takes as input $\bilinear$
    output by $\gen(k)$, $X \in \GGa$ and $(\Pb,Y,Z) \in \GGb^2$ and
    tries to output $(R_1,R_2) \in \GGa \times \GGb$ such that
    $\couple{R_1}{R_2} = \expocouple{xyz}{\Pa}{\Pb}$, where $\Pa =
    \psimorph(\Pb)$, $X = \expo{x}{\Pa}$, $Y = \expo{y}{\Pb}$ and $Z =
    \expo{z}{\Pb}$.  Our exact security reduction relies on two clever
    techniques from \cite{Coron2002a,OgataKurosawaHeng2006}:
    \begin{itemize}
    \item Following a well-known technique due to Coron
      \cite{Coron2002a}, a random coin with expected value $\lambda
      \in [0,1]$ decides whether $\AdverB$ introduces the challenge in
      the answer to the random oracle or an element with a known
      preimage. For the optimal value of $\lambda$, this introduce the
      (small) loss factor $(1 + 6 \cdot \qSig \cdot 2^{f_r})$ in the
      success probability.
    \item Using an approach due to Ogata, Kurosawa and Heng
      \cite{OgataKurosawaHeng2006}, introduced to analyze the security
      of Chaum's undeniable signatures, we do not need a decisional
      oracle to simulate the verification queries. The idea is that,
      unless \blsudvs{} is not unforgeable, all verification queries
      necessarily involve designated verifier signatures that were
      obtained from signing oracles (and can be readily checked) or
      that are invalid. $\AdverB$'s strategy is to guess which
      verification query involves a forged signature and reject
      signatures involved in all other queries. This is done at the
      expense of losing the factor $(\qCon+1)$ in $\AdverB$'s
      probability of success.
    \end{itemize}

    For the ease of presentation, let us (at first) assume that
    $\AdverB$ has an access to a decisional oracle for the problem
    \dsfflex{} (following Okamoto-Pointcheval's so called gap-problems
    \cite{OkamotoPointcheval2001}).
    \begin{Jeux} \item $\AdverB_1$ plays the role of the challenger in
      the experiment \Exper{\schema}{\Adver}{\UDVSEFCMA} of the
      definition \ref{def:UDVS-EFCMA}, in the random oracle model.  It
      plugs the bilinear structure underlying its problem instance in
      the public parameters $\params$. $\AdverB_1$ simulate the random
      oracle $\orHash$ by storing the queries made by $\Adver$ into a
      list denoted $\HList$ (which contains at most $(\qHash(k) +
      \qSig(k) + \qCon(k) + 1)$ 4-tuples). $\AdverB_1$ manage a
      counter $c$ (with initial value $0$) and for each signing,
      verifying or hashing query, $\AdverB_1$ executes the routing
      \textsf{Message}.

      \begin{small}
        \begin{list}{}{}\label{rout:message}
        \item[\textsf{Initialization}] $k$ \\
          $c \leftarrow 0$ \\
          $\params \leftarrow [\bilinear, \Pa, \Pb, n_r]$, \\
          $(u_a, u_b) \hasard \range{q-1}^2$ \\
          $(U_a, U_b) \leftarrow (\expo{u_a}{\Pb},\expo{u_b}{\Pb})$ \\
          \textbf{run} $\Adver(\params,U_a,U_b)$ $\leadsto (\mst,
          \dsigst)$.

        \item[\textsf{Message}] $m$ \\
          \textbf{if} $\exists i \in \range{c}$, $m = m_i$ \\
          \phantom{----} \textbf{then return} $i$ \\
          \phantom{----} \textbf{else} $c \leftarrow c+1$ \\
          \phantom{---- \textbf{else}} $m_c \leftarrow m$ ;
          $\mathcal{L}_c \leftarrow \eps$ ; $\nu
          \hasard [0, 1]$ \\
          \phantom{---- \textbf{else}} \textbf{while} $\nu < \lambda$
          \textbf{and} $\# \mathcal{L}_c \leq \qSig(k)$
          \textbf{do} $\rho \hasard \{0, 1\}^{n_r}$, \\
          \phantom{---- \textbf{else} \textbf{while} $\nu < \lambda$
            \textbf{and} $\# \mathcal{L}_c \leq \qSig(k)$ \textbf{do}}
          $\mathcal{L}_c \leftarrow \mathcal{L}_c \cup \{ \rho
          \}$, \\
          \phantom{---- \textbf{else} \textbf{while} $\nu < \lambda$
            \textbf{and} $\# \mathcal{L}_c \leq \qSig(k)$ \textbf{do}}
          $\nu \hasard [0, 1]$ \\
          \phantom{---- \textbf{else}} \textbf{return} $c$.
        \end{list}
      \end{small}

      The oracle queries are then simulated by $\AdverB_1$ in a
      classical way:
      \begin{small} \begin{list}{}{}
        \item [\textsf{Simulation of the oracles}] \quad \\
          \itemain{} $\orHash(m, r)$: \textbf{if} $\exists R,  (m, r, R,  ?) \in \HList$ \\
          \phantom{\itemain{} $\orHash(m, r)$: \hspace{1cm}} \textbf{then} \textbf{return} $R$\\
          \phantom{\itemain{} $\orHash(m, r)$: \hspace{1cm}} \textbf{else} $\alpha \hasard \range{q-1}$, $i \leftarrow \textsf{Message}(m)$ \\
          \phantom{\itemain{} $\orHash(m, r)$: \hspace{1cm}
            \textbf{else}} \textbf{if} $r \in \mathcal{L}_i$
          \textbf{then} $R \leftarrow \expo{\alpha}{\Pa}$
          \textbf{else} $R \leftarrow \expo{\alpha}{X}$ \\
          \phantom{\itemain{} $\orHash(m, r)$: \hspace{1cm}
            \textbf{else}} $\HList \leftarrow
          \HList \cup \{(m_i, r, R, \alpha)\}$, \\
          \phantom{\itemain{} $\orHash(m, r)$: \hspace{1cm}  \textbf{else}} \textbf{return} $R$ \\
          \itemain{} $\orSig(m)$: $i \leftarrow \textsf{Message}(m)$  \\
          \phantom{\itemain{} $\orSig(m)$:} \textbf{if} $\mathcal{L}_i = \emptyset$ \textbf{then return} $\perp$ \\
          \phantom{\itemain{} $\orSig(m)$: \textbf{if} $\mathcal{L}_i
            = \emptyset$} \textbf{else}
          $r \hasard \mathcal{L}_i$ ; $\mathcal{L}_i \leftarrow \mathcal{L}_i \setminus \{r\}$ \\
          \phantom{\itemain{} $\orSig(m)$: \textbf{if} $\mathcal{L}_i
            = \emptyset$ \textbf{else}}
          $\orHash(m, r)$ ; \textbf{find} $(m_i, r, R, \alpha)$ \textbf{in} $\HList$ \\
          \phantom{\itemain{} $\orSig(m)$: \textbf{if} $\mathcal{L}_i
            = \emptyset$ \textbf{else}}
          $S \leftarrow \expo{\alpha}{U_a}$ \\
          \phantom{\itemain{} $\orSig(m)$: \textbf{if} $\mathcal{L}_i
            = \emptyset$ \textbf{else}}
          \textbf{return} $(r,S)$ \\
          \itemain{} $\orCon(m, \dsig)$:  $i \leftarrow \textsf{Message}(m)$, \\
          \phantom{\itemain{} $\orCon(m, \dsig)$:} $(r,Q_1,Q_2) \leftarrow \dsig$ \\
          \phantom{\itemain{} $\orCon(m, \dsig)$:} $\orHash(m, r)$ ; \textbf{find} $(m_i, r, R, \alpha)$ \textbf{in} $\HList$ \\
          \phantom{\itemain{} $\orCon(m, \dsig)$:} \textbf{return}
          $\dsfflex(R,U_a,U_b,Q_1,Q_2)$.
        \end{list}
      \end{small}

      In the random experiment $\Exp_i$, for $i \in \{1,2\}$, let us
      denote $\mathsf{F_i}$, the event
      \begin{center} ``$\mst \notin \mathcal{Q}_{\orSig}$ and
        $\schema.\dverify(\params,\mst,\pks,\dsigst,\pkc,
        \skc)=\valid$.''
      \end{center}

      Compared to the definition \ref{def:UDVS-EFCMA}, the
      distribution of $\Adver$'s entries is unchanged and the
      simulation of the oracles $\orHash$ and $\orCon$ is perfect.
      Moreover, $\AdverB_1$ answers without aborting to all signature
      queries with probability $\lambda^{\qSig(k)}$, and therefore we
      have
$$\Pr[\mathsf{F_1}] \geq \lambda^{\qSig(k)} \Succ{\schema}{\Adver}{\UDVSEFCMA}(k).$$

\item $\AdverB_2$ replace in the following the public keys $U_a$ and
  $U_b$ furnished to $\Adver$ by the values $Y$ and $Z$ of unknown
  discrete logarithms (in base \Pb).

  \begin{small}
    \begin{list}{}{}
    \item[\textsf{Initialization}] $k$ \\%
      $\params \leftarrow [\bilinear, \Pb, n_r]$, \\
      $(U_a, U_b) \leftarrow (Y,Z)$ \\
      \textbf{run} $\Adver(\params,U_a,U_b)$ $\leadsto (\mst,
      \dsigst)$.
    \end{list}
  \end{small}

  When $\Adver$ returns the pair $(\mst, \dsigst)$, with $\dsigst =
  (r^{\star}, Q_1^{\star}, Q_2^{\star})$, the algorithm $\AdverB_2$
  executes a hash query $\orHash(\mst,r^{\star})$ and gets $i \in \llb
  0, c \rrb$ such that $\mst = m_i$ and $(\mst, r^{\star}, R, \alpha)$
  in $\HList$.
  The algorithm $\AdverB_2$ aborts its execution (by returning
  $\perp$) if $r^{\star} \in \mathcal{L}_i$. Otherwise, $\AdverB_2$
  returns the pair $(\expo{\alpha^{-1}}{Q_1^{\star}},Q_2^{\star})$.

  \medskip
  \begin{small}
    \begin{list}{}{}
    \item[\textsf{End of $\AdverB = \AdverB_5$'s execution}]
      $(\mst,\dsigst)$ \\%
      $(r^{\star},Q_1^{\star},Q_2^{\star}) \leftarrow \dsigst$ \\
      $i \leftarrow \textsf{Message}(m)$ \\
      $\orHash(m, r)$ \\
      \textbf{find} $(m_i, r, R, \alpha)$ \textbf{in} $\HList$ \\
      \textbf{if} $r^{\star} \in \mathcal{L}_i$ \textbf{then return} $\perp$ \\
      \phantom{\textbf{if} $r^{\star} \in \mathcal{L}_i$} \textbf{else
        return} $(\expo{\alpha^{-1}}{Q_1^{\star}},Q_2^{\star})$. \\
    \end{list}
  \end{small}
\end{Jeux}

The probability that $r^{\star} \notin \mathcal{L}_i$ is independent
from $i$ and equal to
$$F(\lambda) = [\lambda (1 - 2^{-n_r})]^{\qSig(k)} + (1-\lambda) \displaystyle\sum_{j = 1}^{\qSig(k)-1} [\lambda(1 -
2^{-n_r})]^{j}.
$$
By the simulation, if $r^{\star} \notin \mathcal{L}_i$ and if the
event $\mathsf{F_2}$ holds, we have $R = \expo{\alpha}{X}$ and if
$\dsigst$ is a valid forgery $\couple{Q_1^{\star}}{Q_2^{\star}} =
\couple{P_1}{P_2}^{xyz\alpha^{\star}}$.  Therefore, the pair
$(\expo{\alpha^{-1}}{Q_1^{\star}},Q_2^{\star})$ is the solution of the
problem \flex{} instance.

The security analysis shows that $\AdverB=\AdverB_2$ satisfies
$$
\Succ{\gen}{\AdverB}{\probl}(k) \geq \lambda^{\qSig(k)} F(\lambda)
\Succ{\schema}{\Adver}{\UDVSEFCMA}(k)
$$
and runs in time at most $\tau' \leq \tau + (\qHash + \qSig + \qCon +
1)(\Texp{\GGa} + O(1)) + \Texp{\GGc} $ while making at most $\qCon(k)$
queries to the decisional oracle \dsfflex. An easy computation gives
proves the existence of a value $\lambda_0$ such that
$\lambda_0^{\qSig(k)} F(\lambda_0) \geq (1 + 6 \cdot \qSig \cdot
2^{f_r})$ (see \cite{Coron2002a}).

\renewcommand{\probl}{\flex} In this reduction, if the decisional
oracle for the problem \dsfflex{} returns \valid{} for a $5$-tuple
$(R, Y, Z, Q_1, Q_2)$ associated to a verifying query on a pair $(m_i,
(r, Q_1, Q_2))$ and if $r \notin \mathcal{L}_i$, then the pair
$(\expo{\alpha^{-1}}{Q_1},Q_2)$ is a solution of the problem \flex{}
and there is no need to continue the execution of $\Adver$. By using
this remark, it is possible to prove the resistance to forgery of the
scheme to the problem \probl{} without using the decisional oracle.

\medskip A verifying query (made by $\Adver$ or $\AdverB$ at the end
of its execution) on a pair $(m_i, (r, Q_1,Q_2))$ is said
\emph{special} if $r$ does not belong to the list $\mathcal{L}_i$. Let
us denote $\mathsf{A}$ the event: ``One special verifying query is
made in the random experiment $\Exp_5$'', $\mathsf{A_i}$ the event
``The first special verifying query in the experiment $\Exp_5$ is the
$i$-th'', for $i \in \range{\qCon(k)}$ and $A_{\qCon(k)+1}$ the event
``The first special verifying query in the experiment $\Exp_5$ is the
one on the pair $(\mst, \dsigst)$''. We have
$$
\mathsf{A} = \displaystyle\bigsqcup_{i = 1}^{\qCon(k)+1} \mathsf{A_i}
\hbox{ et } \widetilde{\mathsf{F_2}} \subseteq \mathsf{A}.
$$
where $\widetilde{\mathsf{F_2}}$ is the event
``$\schema.\dverify(\params,\mst,\pks,\dsigst,\pkc,\skc)=\valid$,
$\mst=m_i \notin \mathcal{Q}_{\orSig}$ and $r^{\star} \notin
\mathcal{L}_i$.''

In this variant, the algorithm $\AdverB$ picks uniformly at random an
integer $v \in \range{\qCon(k)+1}$ at the beginning of its execution.
For each verifying queries on $(m_i, \dsig)$ where $\dsig =(r, Q_1,
Q_2)$, $\AdverB$ gets the $4$-tuple $(m_i, r, R, \alpha)$
corresponding to the hash value of $(m_i, r)$ and
\begin{itemize}
\item if $r \in \mathcal{L}_i$ then $\AdverB$ returns $\valid$ if and
  only if $\couple{Q_1}{Q_2} =
  {\couple{\psimorph(U_a)}{U_b}}^{\alpha}$;
\item if $r \notin \mathcal{L}_i$ and if the verifying query is not
  the $v$-th, then $\AdverB$ returns $\invalid$;
\item if $r \notin \mathcal{L}_i$ and if the verifying query is the
  $v$-th, $\AdverB$ stop $\Adver$'s execution and returns the pair
  $(\expo{\alpha}{Q_1},Q_2)$.
\end{itemize}

\medskip If the event $\textsf{A}_v$ occurs then the simulation of the
oracles done by $\AdverB$ until the $v$-th verifying query is
indistinguishable from the previous one and $(\expo{\alpha}{Q_1},Q_2)$
is actually the solution to the instance $(P_2, X, Y, Z)$ of the
problem \probl{}.  Consequently, we have
\begin{eqnarray*}
  \Succ{\gen}{\AdverB}{\probl}(k)  \geq \sum_{i = 1}^{\qCon(k)+1} \Pr[\mathsf{A_i}] \cdot \Pr[i = v] 
  & = & \displaystyle{\frac{1}{\qCon(k)+1} \sum_{i = 1}^{\qCon(k)+1}
    \Pr[\mathsf{A_i}]} \\
  & = & \frac{\Pr[\mathsf{A}]}{\qCon(k)+1} \\
  & \geq & \frac{\Pr[\widetilde{\mathsf{F_2}}]}{\qCon(k)+1}.
\end{eqnarray*}

This variant of $\AdverB$ does not make any call to the oracle \dbdh{}
and its execution time is increased by at most one exponentiation in
the group $\GG_3$ by each query to the oracle \orCon{}. This gives the
claimed result.

\item The proof is more or less routine (see
  \cite{LaguillaumieVergnaud2005a,OgataKurosawaHeng2006} for instance)
  and therefore left to the reader.
\end{enumerate}}

\bigskip \remark{If the public verification is desirable in an
  application (\eg{} to design a UMDVS scheme) or if the anonymity
  property is not necessary, the unforgeability of the protocol
  \blsudvs{} can be reinforced. It is indeed possible to add a fourth
  element to the signature (namely $Q_3 = \expo{t^{-1}}{P_2} \in \GGb$
  allowing the public verification) in such a way that the scheme
  obtained is really close to the protocol \bbudvs. Since a designated
  verifier signature for the protocol \blsudvs{} can be readily
  derived from one for this scheme and since the underlying signature
  scheme is the same, we get immediately that forging a signature for
  the latter scheme is at least as hard as for \blsudvs. Under the
  knowledge-of-exponent-assumption in $\GGb$, we can also prove, in
  the standard security model, the resistance to forgery of the scheme
  assuming only the \EFCMA-security of the underlying signature scheme
  \bls{}.}

\section{Extension of the schemes to the Multi-verifier setting}\label{app:multi}%
At Crypto'03 rump session, Desmedt opened the question to allow
several designated verifiers in designated verifier signatures. The
first step towards this problem was made in
\cite{LaguillaumieVergnaud2004} with the introduction of the
\emph{multi designated verifiers signature} primitive and some
concrete realizations of it. The notion of \emph{universal multi
  designated verifier signatures} was naturally proposed shortly
afterwards in \cite{NgSusiloMu2005}.

\subsection{\bbudvs}
Let $n \in \NN$. The scheme $\bbudvs$ can be seen as a ``discrete-log
two-party ring signatures'' and therefore, following the generic
construction from \cite{LaguillaumieVergnaud2004}, it can readily be
extended into a universal $n$-designated verifier signature schemes:
the algorithm $\vkeygen$ remains unchanged and in the signing
algorithm, the verifying public key $\pkV$ is simply replaced by the
sum of the $n$ verifying public keys
$$\pkV_1 +
\cdots + \pkV_n = \expo{\skV_1 + \cdots + \skV_n}{P_2}.$$ Using a
multi-party computation (for instance) and the algorithm
$\bbudvs.\fake$, the designated verifiers can cooperate to produce an
$n$-designated verifier signature from $\pkS$ to the keys $(\pkV_1, \ldots,
\pkV_n)$. This fact, with the source hiding property of $\bbudvs$
ensure the same property for the multi-user protocol.  Finally, since
the algorithm $\bbudvs.\dverify$ is public (\ie\ does not require the
verifying secret key) the algorithm $\dverify$ is identical in the
multi-user setting with the verifying public key $\pkV$ replaced again
by the sum of the $n$ verifying public keys $\pkV_1 + \cdots +
\pkV_n$. In particular, it is very efficient and does not require
interaction between the designated verifiers. It is worth noting, that
in order to avoid well-known \emph{rogue key attacks}, the users
should prove the knowledge of their secret key (in the registered
public key model, for instance).

\subsection{\blsudvs}
The scheme \blsudvs{} is not publicly verifiable and therefore it does
not enter in the generic construction proposed in
\cite{LaguillaumieVergnaud2004}. However, it is possible to adapt this
scheme in order to design a universal $n$-designated verifier
signature scheme for all integer $n \geq 1$. With the previous
notations, suppose that Alice (\resp{} the $n$ verifiers) has
published a public key $U_a = \expo{u_a}{P_2}$ (\resp{} $U_{b_i} =
\expo{u_{b_i}}{P_2}$ for $i \in \range{n}$) and that the pair $\sig =
(r, S) \in \range{q-1} \times \GG_1$ is a signature \bls{} produced by
Alice on a message $m$. If Cindy wants to designate $\sig$ to the set
of the $n$ verifiers, she picks uniformly at random an integer $t \in
\range{q-1}$ and sets $Q_0 = \expo{t}{S}$ and for all $i \in
\range{n}$, $Q_i = \expo{t^{-1}}{U_{b_i}}$. The $(n+1)$-tuple $\dsig =
(r, Q_0, Q_1, \ldots, Q_n)$ is the multi-designated verifier signature
on $m$.

\medskip The pairing insures the correctness of the scheme since the
$i$-th verifier can check the consistency of the multi-DVS by checking
for all $j \in \range{n} \setminus \{i\}$ if the equality $
\couple{\psimorph(U_j)}{Q_i} = \couple{\psimorph(U_i)}{Q_j}, $ holds
and then ascertain its validity thanks to its knowledge of its secret
key by verifying the equality: $\couple{Q_0}{Q_i} =
\couple{\expo{u_{b_i}}{\Hash(m,r)}}{U_a}$.

\medskip The security properties of the scheme \blsumdvs{} are similar
to those of the scheme \blsudvs{}. In the security reduction of
unforgeability, a factor $1/n$ is lost. This factor corresponds to the
bet made by the algorithm \AdverB{} on the public key that will not
corrupt the adversary $\Adver$. Once this choice has been made, the
proof is identical to the one of the theorem \ref{the:BLSUDVS} and we
can easily prove the unforgeability and the anonymity of this scheme
assuming the intractability in \gen{} of the problems \flex{} and
\dsfflex{} (respectively).

\section*{Acknowledgements} It is a pleasure to acknowledge Fabien
Laguillaumie and Beno\^{\i}t Libert for their great comments and
simplifying suggestions on a preliminary version of this paper. I am
also grateful to Willy Susilo and Rui Zhang for providing a copy of
their papers \cite{NgSusiloMu2005,ZhangFurukawaImai2005}. Finally, I
would like to thank the referee for its careful reading of this paper. 

\providecommand{\bysame}{\leavevmode\hbox to3em{\hrulefill}\thinspace}
\providecommand{\MR}{\relax\ifhmode\unskip\space\fi MR }
\providecommand{\MRhref}[2]{%
  \href{http://www.ams.org/mathscinet-getitem?mr=#1}{#2}
}
\providecommand{\href}[2]{#2}

\end{document}